\documentclass{article}
\usepackage{amsmath}
\usepackage[latin1]{inputenc}
\usepackage{multicol}
\usepackage{graphicx}
\usepackage{psfrag}
\newcommand{\al}{\alpha}
\newcommand{\beq}{\begin{equation}}
\newcommand{\eeq}{\end{equation}}
\newcommand{\bfg}{\begin{figure}}
\newcommand{\efg}{\end{figure}}

\def\eq#1{(\ref{#1})}
\def\fig#1{Fig.~\ref{#1}}
\def\bgrk#1{\mbox{{\boldmath $#1$ \unboldmath}}\!\!}
\def\bea{\begin{eqnarray}}
\def\eea{\end{eqnarray}}

\begin{document}

\title{Effect of pitchfork bifurcations \\  on the spectral statistics of
        Hamiltonian systems }
\author{Marta Gutiérrez, Matthias Brack, and Klaus Richter\\
{\small \it Institute for Theoretical Physics, University of Regensburg, Germany}\\
\\
Ayumu Sugita\\
{\small \it Osaka City University, Osaka, Japan}}
 
\maketitle

\begin{abstract}
We present a quantitative semiclassical treatment of the effects
of bifurcations on the spectral rigidity and the spectral form factor
of a Hamiltonian quantum system defined by two coupled quartic 
oscillators, which on the classical level exhibits mixed phase space 
dynamics.  
We show that the signature of a pitchfork bifurcation is two-fold:
Beside the known effect of an enhanced periodic orbit contribution
due to its peculiar $\hbar$-dependence at the bifurcation, we 
demonstrate that the orbit pair born {\em at} the bifurcation gives
rise to distinct deviations from universality slightly {\em above} the 
bifurcation. This requires a semiclassical treatment beyond the so-called
diagonal approximation. Our semiclassical predictions for both the 
coarse-grained density of states and the spectral rigidity, are in
excellent agreement with corresponding quantum-mechanical results.
\end{abstract}

\maketitle

\newpage

\section{Introduction}

A prominent approach to the quest of ``quantum chaos'' involves
spectral statistics to characterize the energy-level fluctuations in 
quantum systems and their interpretation in terms of the dynamics of the 
corresponding classical system. 
Classically integrable systems possess uncorrelated energy levels,
described by a Poisson distribution \cite{ref:BeTa}, while the levels 
of classically chaotic quantum systems exhibit strong local repulsion.
This behaviour is conjectured to be the same as for the
eigenvalues of ensembles of random matrices preserving certain
general symmetries \cite{ref:BoGiSch}. Spectral statistics 
has been investigated, for both integrable 
\cite{ref:CasChi,ref:RobVe,ref:Bog} and chaotic 
\cite{ref:HaAl,ref:Be85,ref:SiKiSmKea} systems,
employing semiclassical (periodic orbit) approaches, which provide the 
closest link between classical and quantum properties.
For the purely chaotic case, starting with Ref.\ \cite{ref:SiRi},
considerable progress has been recently made in understanding energy 
level correlations semiclassically beyond the so-called diagonal approximation 
\cite{ref:Be85} by means of classical correlations between (off-diagonal pairs)
of periodic orbits \cite{ref:off-diagonal}.
 
However, integrability and full chaoticity represent extreme situations 
which occur rather exceptionally. The most realistic physical 
situation is that of a system which is neither completely chaotic nor 
integrable, but whose phase space contains a mixture of stable orbits
surrounded by regular islands and chaotic regions. One main feature
and structuring element of classical mixed phase space dynamics is the 
occurrence of bifurcations of periodic 
orbits upon variations of the energy or other parameters of the Hamiltonian. 
Moreover bifurcations lead to noticeable effects in the spectral statistics,
because in semiclassical trace formulae for the density of states
\cite{ref:Gutzbook,ref:BraBa}, contributions from periodic orbits at a 
bifurcation exhibit an enhanced weight, compared to that of isolated orbits. 
This has been demonstrated for the generalized cat map in Ref.~\cite{ref:BeKeaPra}, 
where the semiclassical signature of a tangent bifurcation was studied 
on the level of the  diagonal approximation. 


More generally, in Ref.\ \cite{ref:BeKeaScho,ref:KeaPraSie} a semiclassical 
approach was developed for the moments of the level counting function in the 
presence of several competing generic bifurcations. It was suggested that 
these moments diverge with a universal ``twinkling exponent'' in the 
semiclassical limit $\hbar \to0$. 

In the present paper we investigate the role of pitchfork bifurcations
on the spectral statistics in Hamiltonian systems that are closer to a 
realistic physical situation than the maps considered so far. 
We show that bifurcations of short orbits has
a considerable effect on the spectral rigidity and the spectral form factor,
respectively, even in the almost chaotic case.
As a standard system with mixed classical dynamics, we choose the
Hamiltonian 
of two coupled quartic oscillators. Its relevant classical bifurcation 
characteristics is summarized in Sec.~\ref{sec2}. In Sec.~\ref{sec3}
we present a detailed semiclassical analysis including a comparison
with quantum results for the (smoothed) density of states for different
symmetry classes, as a prerequisite for the treatment of spectral correlations 
in Sec.~\ref{secrig}. There we quantitatively analyze
deviations of the spectral rigidity from universality employing 
uniform approximations to derive the semiclassical periodic orbit weights at
the bifurcation. We show, in particular, that pairs of orbits 
(with an action difference smaller than Planck's constant $\hbar$),
born at a pitchfork bifurcation, yield important non-diagonal contributions 
to the spectral form factor and rigidity. The deviations from the quantum 
chaotic universality are found to be most significant {\em after},
rather than {\em at} the bifurcation.
 
\section{The quartic oscillator Hamiltonian}
\label{sec2}

As a representative system 
we investigate the coupled quartic 
oscillator (QO) in two dimensions. Its Hamiltonian reads:
\beq\label{QO}
H(x,y,p_x,p_y)=\frac{1}{2}\,(p_x^2+p_y^2)+\frac{1}{4}\,(x^4+y^4)
               +\frac{\al}{2}\,x^2y^2 \, .
\eeq
It has been extensively studied both classically, semiclassically 
and quantum-mechanically
\cite{ref:Boh93,ref:Eck,ref:BraMehTa,ref:BraFeMagMeh,ref:erda}, 
as a smooth potential
model exhibiting the transition from integrability to 
chaotic behaviour. 
Here we summarize the main classical features relevant for the
subsequent semiclassical treatment.
Since the Hamiltonian \eq{QO} is homogeneous, its classical dynamics at 
different energies $E$ can be related to each other by a simple scaling 
of coordinates, momenta and time. All actions scale with energy $E$ as
$E^{3/4}$ so that the semiclassical limit can be unambiguously taken
as $E\to\infty$. 

After scaling out the energy the parameter $\alpha$ in Eq. \eq{QO} 
solely determines the dynamics. The system 
is integrable for $\al=0$, 1, and 3. For $\alpha=9$, it is almost 
completely chaotic: we could not locate any 
stable periodic orbit with a period up to about four times that of the
shortest orbits. For values $\alpha>9$ the regular fraction of the 
phase space keeps oscillating with a decreasing amplitude. 

The QO in Eq. \eq{QO} possesses periodic straight-line librational 
orbits along 
both axes which we label by A. They undergo stability oscillations 
under the variation of $\al$. Infinite cascades of new periodic 
orbits bifurcate from the A orbits and their repetitions. The motion 
of the A libration can be given analytically in terms of Lamé
functions \cite{ref:BraMehTa,ref:BraFeMagMeh}. The trace of its 
stability matrix M (see \cite{ref:Gutzbook,ref:BraBa} for its
definition)  as a function of $\alpha$ is known analytically 
\cite{ref:Yos}: 
\beq\label{trA}
\rm{Tr\,M}(\alpha)=4\cos \left(\frac{\pi}{2}\sqrt{1+8\al} \right)+2\,.
\eeq
Isochronous pitchfork bifurcations of the A orbit (which are non-generic 
due to the discrete symmetries of the system) take place when
Tr\,M=+2, i.e., for
\beq\label{an}
\al=\al_n=\frac{1}{2}\,n\,(n+1)\,, \qquad \qquad n=0,3,4,5,\dots \;\; .
\eeq
(For $\alpha_1=1$ and $\alpha_2=3$, where the system is integrable,
the A orbit is member of a degenerate family and does not bifurcate.
See also \cite{ref:BraMehTa,ref:BraFeMagMeh} for more details about the
periodic orbits of this system.)

\begin{figure}[htbp]
\centerline{\includegraphics[width=12.5cm]{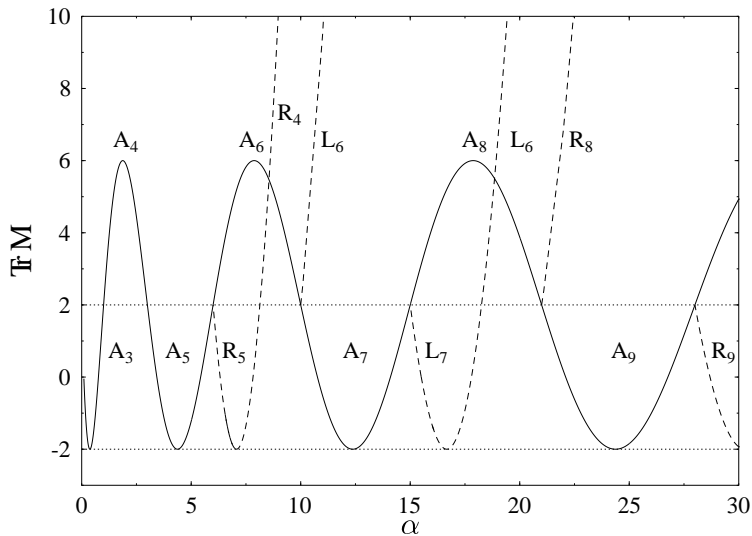}}
\caption{Trace of the stability matrix M as a function of $\alpha$,
Eq.\ \eq{QO}, for the primitive A orbit 
(solid line) and the new orbits born at its bifurcations 
(dashed lines) at $\alpha=6$, 10, 15, 21, and 28.
Subscripts denote the Maslov indices $\sigma_j$ (see Sec.~\ref{sec3}).
}\label{fig:traMa}
\end{figure}

In \fig{fig:traMa} we show Tr\,M$(\al)$ for the primitive A orbit and 
the new orbits born at its bifurcations at $\al_n$ with $n=3$ to 7.
These orbits are alternatingly stable or unstable rotational (R$_\sigma$) 
and librational orbits (L$_\sigma$) with a classical degeneracy of 2 
due to the symmetries (cf.\ Ref.\ \cite{ref:BraMehTa}.) In
our numerical case studies below, we shall focus on the bifurcation at
$\al=\al_4=10$ where the orbit L$_6$ is born. Note that at each second 
bifurcation ($n=3,5,\dots$) a new stable orbit (R$_5$, L$_7,\dots$) is 
born, so that stable orbits exist on either side of these bifurcations. 
At the other bifurcations ($n=4,6,\dots$), on the other hand, the new 
orbits (L$_6$, R$_8,\dots$) are unstable, and just before these 
bifurcations, the A orbit is also unstable. This explains the oscillating  
regularity of the phase space and the fact that, even in the limit 
$\alpha \to\infty$, there always exist regions with stable orbits 
\cite{ref:erda}).

The potential in Eq.\ \eq{QO} is invariant under the symmetry operations 
that conform the point group symmetry $C_{4V}$, which has four 
one-dimensional irreducible representations and one (doubly-degenerate) 
two-dimensional representation. Due to the $C_{4V}$ symmetry, the
full eigenvalue spectrum would not exhibit any universal statistics.
For an appropriate study of the spectral statistics, each symmetry class 
must therefore be treated separately. We shall study mainly the 
representation corresponding to eigenfunctions which are symmetric 
under the operations $x\to -x$, $y\to -y$ and $x\to y$, which we call 
EES. This representation is easier to handle semiclassically, because all 
its characters are equal to unity.

For the numerical calculation of the quantum mechanical eigenenergies
we follow the procedure outlined in Ref.\ \cite{ref:PuEd}. 
We diagonalise the Hamiltonian 
using a basis of symmetry-adapted linear combinations of harmonic 
oscillator states:
\beq |n_x,n_y\rangle_m=\frac{1}{\sqrt{2}}(|n_1,n_2\rangle 
   ± |n_2,n_1\rangle)\,,
\eeq
where the sign and the parity of $n_1$ and $n_2$ depends on the 
representation. Since the independent symmetry-reduced blocks of the 
Hamiltonian matrix in this basis are banded, we can obtain up to 
tenthousand well-converged eigenvalues, allowing for significant statistics.

\section{Semiclassical density of states for discrete symmetries}
\label{sec3}

Periodic orbit theory yields the semiclassical spectral density as
\beq  \label{POT}
g(E)=\bar d(E)+\delta g(E)
\eeq
where the smooth part $\bar d(E)$ is given by the (extended)
Thomas-Fermi model (cf.\ Chap. 4 in Ref.\ \cite{ref:BraBa}), and the
oscillating contribution is given by a trace formula which, to leading 
order in $1/\hbar$, has the following form:
\beq \label{dens}
\delta g(E)=\frac{1}{\hbar^{\mu+1}}\sum_j A_j(E) 
            \cos\left[\frac{S_j(E)}{\hbar}-\frac{\pi}{2}\sigma_j\right]\!.
\eeq
The sum is over all periodic orbits $j$ (which form families with
degenerate actions in the presence of 
continuous symmetries).
$S_j(E)=\oint_j {\bf p}\cdot d{\bf q}$ is the action 
integral along a periodic orbit and $\sigma_j$ a geometrical phase factor
(usually called Maslov index). The amplitudes $A_j(E)$ and the
power of $\hbar$ in Eq.\ \eq{dens} depend on the presence of continuous
symmetries. For systems without continuous symmetries, where all
orbits are isolated in phase space, one has $\mu=0$, and the amplitudes
$A_j(E)$ were given by Gutzwiller \cite{ref:Gutz} in terms of
their stability matrices M$_j(E)$ and periods $T_j(E)=$ d$S_j(E)$/d$E$.
When an isolated periodic orbit undergoes a bifurcation at an energy
$E_0$, its amplitude in the Gutzwiller trace formula  
diverges and uniform approximations 
must be developed \cite{ref:Ozoha} to obtain a finite $A_j(E_0)$; in
this case one finds $0 < \mu \leq 1/2$, the precise value of $\mu$ depending 
on the generic type of the bifurcation (cf.\ also Ref.\  \cite{ref:SchoSie}).
For fully integrable systems, $\mu=f/2$, where $f$ is the degree of
degeneracy of the most degenerate orbit families; the amplitudes 
were derived by Strutinsky and Magner \cite{ref:StruMa} for specific
cases and by Berry and Tabor \cite{ref:BeTa76} for general integrable
systems (cf.\ also Sec.\ \ref{sec3int} below). For non-integrable systems
with continuous symmetries, further results were obtained by Creagh and 
Littlejohn \cite{ref:CrLi}, who also derived a Berry-Tabor-like trace
fromula for integrable systems. 

In the presence of discrete symmetries it is necessary to define
partial densities of states corresponding to the subspectra of
each irreducible representation of the symmetry group. For systems
with isolated orbits, the corresponding symmetry-reduced semiclassical 
trace formulae have been derived in Refs.~\cite{ref:Rob,ref:Lau,ref:crsr}; 
we shall discuss and use them in Sec.\ \ref{sec3iso}. 

For practical purposes, it is useful to coarse-grain the density of states by
convolution with a normalized Gaussian  
$\exp[-(E/\gamma)^2]/(\sqrt{\pi}\gamma)$.
Hence, we replace the quantum density of states
$d(E)=\sum_n\delta(E-E_n)$ by the ``coarse-grained'' density of states
\beq\label{qmcg}
d_\gamma(E)=\frac{1}{\sqrt{\pi}\gamma}\sum_n\exp\left[-\frac{(E-E_n)^2}{\gamma^2}\right],
\eeq
whereby the smoothing width $\gamma$ defines the energy resolution
at which one wishes to study the spectrum. The correspondingly averaged
semiclassical level density becomes, to leading order in $\hbar$ 
(see, e.g., Ref.\ \cite{ref:BraBa}),
\beq\label{smcg}
\delta g_\gamma(E)=\frac{1}{\hbar^{\mu+1}}\sum_j A_j(E)
                   \exp\left[-\left(\frac{\gamma T_j(E)}{2\hbar}\right)^2\right]
                   \cos\left[\frac{S_j(E)}{\hbar}-\frac{\pi}{2}\sigma_j\right]\!.
\eeq
Hence, long orbits are exponentially suppressed which avoids
convergence problems for not too small values of $\gamma$.

\subsection{Integrable Systems}\label{sec3int}

For integrable systems with $f$ degrees of freedom, it is useful to 
work with action-angle variables $({\bf I},\bgrk{\phi})$, with each 
set of actions ${\bf I}=\{I_1,\ldots,I_f \} $ defining a phase-space 
torus \cite{zhilinskii}. 
The Hamiltonian can be transformed to $H({\bf I})=E$, and the 
frequencies ${\rm d}\bgrk{\phi} / {\rm d}t =\bgrk{\omega}
=\{\omega_1,\ldots,\omega_f\}$ on the torus ${\bf I}$ are given by 
$\bgrk{\omega}({\bf I})=\bgrk{\nabla} H({\bf I})$.
Assuming smooth boundaries, the Einstein-Brillouin-Keller (EBK)
quantization \cite{ref:KelRub}
\beq\label{ebk}
I_j(n_j)=\hbar(n_j+1/2)\,, \qquad n_j=0,1,2,,\ldots\,,\qquad j=1,\dots,f\,,
\eeq
defines a set of $f$ quantum numbers ${\bf n}=(n_1,\dots,n_f)$.
Upon inserting Eq.\ \eq{ebk} into $E=H({\bf I})$, the EBK spectrum reads
\beq\label{Eebk}
E^{EBK}_{\bf n}=E^{EBK}_{n_1,\dots,n_f}=H(I_1(n_1),\dots,I_f(n_f))\,.
\eeq
Berry and Tabor \cite{ref:BeTa76} started from the density of states in 
terms of the $E^{EBK}_{\bf n}$ and converted it, by means of Poisson
summation, into a semiclassical trace formula of the type of Eq.\ \eq{dens}.

The EBK quantization of the integrable QO, Eq.~\eq{QO}, with $\alpha=0$ has been
performed in Ref.\ \cite{ref:BraFeMagMeh}; we quote here those results which are
relevant for our present application. The EBK spectrum becomes
\beq \label{qoebk}
E_{n_x,n_y}^{EBK}=\frac{1}{4}\left(\frac{3\pi\hbar}{2{\bf
                  K}}\right)^{\!\frac43}
                  \left[\left(n_x+\frac12\right)^{\frac43}+\left(n_y+\frac12\right)^{\frac43}\right]\!,
                  \quad (n_x,n_y=0,1,2,\dots)
\eeq
where ${\bf K}=K(\kappa)$ is the complete elliptic integral of
first kind with modulus $\kappa=1/\!\sqrt{2}$. Since the Hamiltonian
\eq{QO} is separable for $\alpha=0$, we can write $E_{n_x,n_y}^{EBK}=
E_{n_x}^{EBK}+E_{n_y}^{EBK}$. The separate one-dimensional densities
of states, 
\beq \label{1dds}
g_j(E)=\sum_{n_j=0}^{\infty}\delta\left(E-E_{n_j}^{EBK}\right)\,, \qquad (j=x,y)
\eeq
which are identical due to the symmetry, become after Poisson summation
\beq \label{1tf}
g_j(E)=\frac{T_A(E)}{2\pi\hbar}\sum_{k_j=1}^{\infty}(-1)^{k_j}
       \cos[k_jS_A(E)/\hbar]\,, \qquad (j=x,y)
\eeq
corresponding to the Gutzwiller trace formula for a one-dimensional
system. Here 
\beq
S_A(E) = \frac{4}{3} {\bf K}  (4E)^{3/4}\,,
\eeq
 is the action of the primitive A orbit
and $T_A(E)={\rm d}S_A(E)/{\rm d}E$ its period.
The total density of states of the full two-dimensional system can then
be written as a convolution integral of the one-dimensional densities:
\beq\label{fold} 
g(E)=\int_0^Eg_x(E-E')\,g_y(E')\,{\rm d}E'.
\eeq
The asymptotic evaluation \cite{ref:Wong} of this integral in the
limit $\hbar\to 0$ yields for the oscillating part
\bea\label{denQO}
\delta g(E) &\!\! = \!\! &  2 \left(\frac{2{\bf
                    K}}{\pi\hbar}\right)^{\!\!\frac32}\!(4E)^{\frac18}\!
                    \sum_{k_x=1}^{\infty}\sum_{k_y=1}^{\infty}(-1)^{k_x+k_y}
                    \frac{k_xk_y}{(k_x^4+k_y^4)^{\frac58}}
                    \cos\!\left[\!\frac{1}{\hbar}S_{k_xk_y}(E)-\frac{\pi}{4}\right]
                    \nonumber\\ 
            & & +   \frac{(4{\bf
                    K})^{\frac34}}{(\pi\hbar)^{\frac54}}\, (4E)^{-\frac{1}{16}}
                    \sum_{k=1}^{\infty}(-1)^{k}\frac{1}{k^{\frac34}}
                    \cos\!\left[\frac{k}{\hbar}S_A(E)-\frac{3\pi}{8}\right]\!. 
\eea
The double sum in the first line above contains the contributions from the 
standard stationary-phase evaluation of the integral. It corresponds exactly 
to the Berry-Tabor trace formula \cite{ref:BeTa76}, whereby the two numbers
$k_x,k_y$ label the rational tori corresponding to the simply degenerate 
families of periodic orbits with two-dimensional motion.
The actions of these rational tori are given by
\beq
S_{k_xk_y}(E)=S_A(E)(k_x^4+k_y^4)^{1/4}.
\eeq

The term in the second line of Eq.\ \eq{denQO} arises from the  boundaries 
of the integral \eq{fold}, corresponding to the A orbits which are 
one-dimensional librations with all energy in either $x$ ($E'=0$) or $y$ 
direction $(E'=E)$. Note that the amplitude of this term involves a prefactor 
$\hbar^{-5/4}$. This is due to the fact that the A orbit undergoes 
a pitchfork bifurcation at $\alpha=0$ corresponding to $n=0$ in
Eq.\ \eq{an}. (The orbits L$_3$ born at this bifurcation exist only for
$\alpha\leq 0$.) In Ref.~\cite{ref:BraFeMagMeh}, identically the same
result \eq{denQO} was obtained, whereby the local uniform approximation 
\cite{ref:Ozoha} for the contribution of the bifurcating A 
orbit was employed \cite{footnote1}. 

\begin{figure}[htbp]
\psfrag{d_osc(E)}{\large $\delta g(E)$}
\psfrag{E}{$E$}
\psfrag{d_ees(E)}{\large $\delta g^{EES}(E)$}
\begin{center}
\includegraphics[width=12cm]{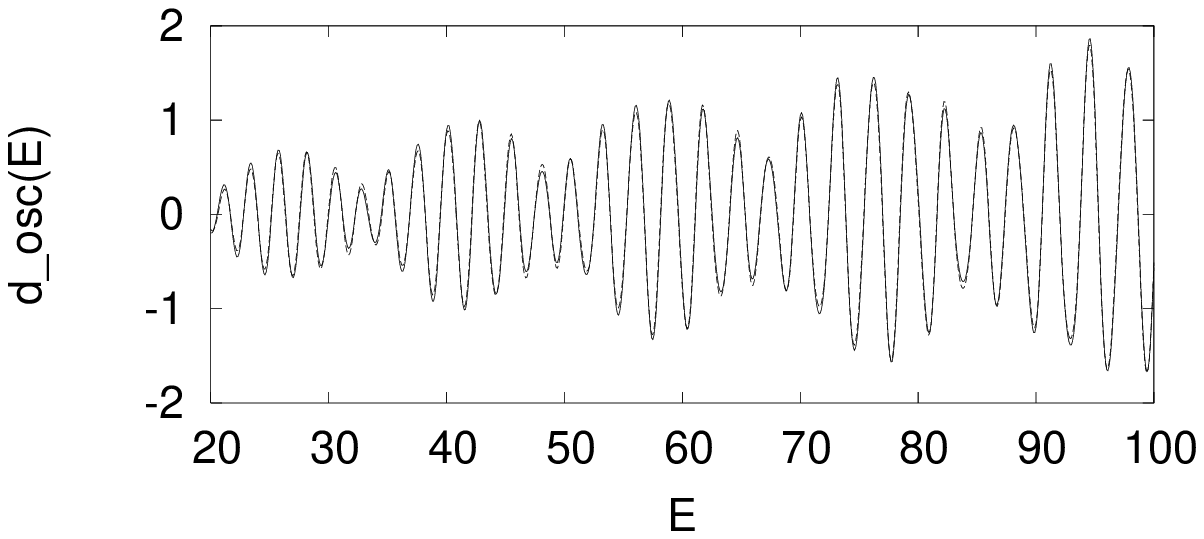}
\includegraphics[width=12cm]{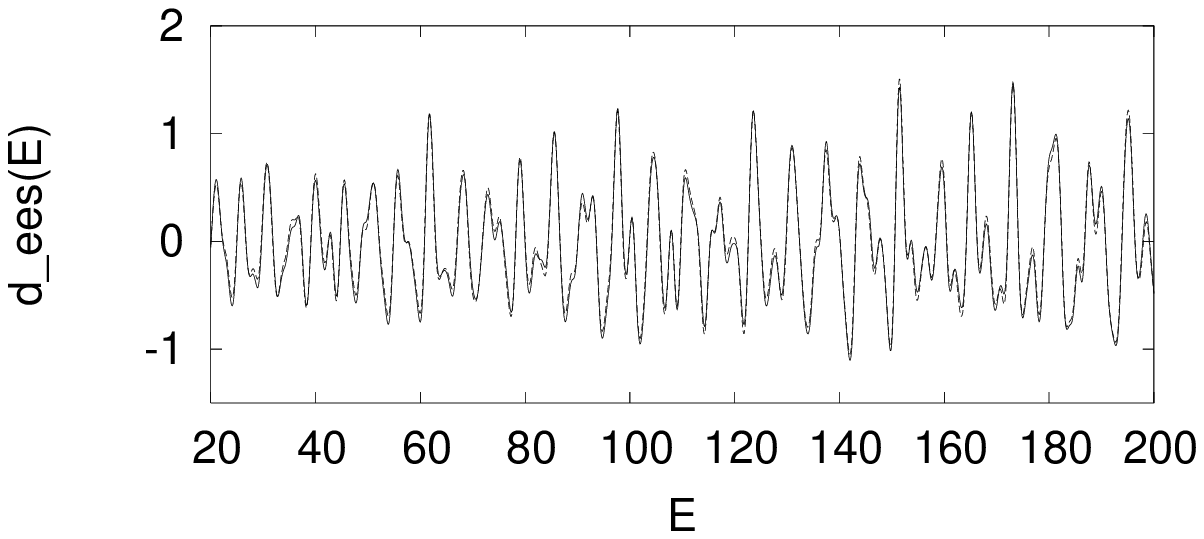}
\end{center}
\caption{Upper panel: total density of states for $\al=0$ coarse-grained by a Gaussian with width $\gamma=1$. Lower panel: symmetry-reduced density of states for the representation
EES, see text.
 Solid line: quantum result, dashed line: semiclassical 
 result, Eq.~\eq{denQO}. 
}
\label{fig:dtot}
\end{figure}

In the upper panel of \fig{fig:dtot} we compare the semiclassical
density of states, Eq.~\eq{denQO} (dashed line), with the 
corresponding quantum-mechanical one (solid line), 
both coarse-grained with a Gaussian average with width 
$\gamma=1$. We find perfect agreement up to very high energies.

We now calculate the symmetry-reduced densities of states by restricting 
ourselves to the subspectra, $E^{EBK}_{\bf n}$, of a given irreducible 
representation. Hereby we can relate the parities of the quantum numbers 
to the symmetries of the irreducible representations.
Thus, we restrict $n_x$ and $n_y$ to be even or odd, according to a 
given representation. For example, let us take the one-dimensional
irreducible representation EES. This corresponds to taking $n_y\leq n_x$ 
with $n_x,n_y$ even. Then the partial density of states  can be calculated
as a convolution 
\begin{equation*}
\delta g^{EES}(E)=\int_0^Eg_x^{E}(E-E')\,g_y^{E}(E')\,{\rm d}E'
\end{equation*}
of the one-dimensional densities $g_j^E(E)$
defined as in Eq.~(\ref{1dds}), except that only 
the terms with even $n_j$ are included in the sum. The asymptotic
evaluation of the convolution integral leads to
\bea\label{rdei}
\delta g^{EES}(E)&\!\!\!\!\!=\!\!\!\!\!&\!\!\!\left(\frac{{\bf K}}{\pi\hbar}
                  \right)^{\!\!\frac32}\!\!(4E)^{\frac18}\!\!\!\!
                  \sum_{k_x,k_y=1}^{\infty}\!
                  \frac{k_xk_y}{(k_x^4+k_y^4)^{5/8}}
                  \cos\!\left[\!\frac{1}{2\hbar}S_{k_xk_y}(E)\!-\!\frac{\pi}{2}(k_x\!+\!k_y)\!
                              -\!\frac{\pi}{4}\right]\nonumber\\
      & & +\frac{1}{2^{\frac34}}\frac{({\bf K})^{\frac34}}{(\pi\hbar)^{\frac54}}
                  \,(4E)^{-\frac{1}{16}}\sum_{k=1}^{\infty}\,
                  \frac{1}{k^{\frac34}}\cos\!\left[\frac{k}{2\hbar}S_A(E)
                  -\frac{\pi}{2}k-\frac{3\pi}{8}\right]\!.
\eea
Again, the first term above corresponds to the Berry-Tabor result for
the rational tori, and the second term comes from the bifurcating A orbit.

In the lower panel of \fig{fig:dtot} we compare the semiclassical and
quantum-mechanical density of states, $\delta g^{EES}(E)$, coarse-grained 
with a Gaussian average with width $\gamma=1$. Again the agreement 
is nearly perfect.

\subsection{Isolated orbits}\label{sec3iso}

The symmetry-reduced densities of states for isolated orbits have been 
derived in Ref.\ \cite{ref:Rob,ref:Lau} by projecting the semiclassical
Green function onto the irreducible representations and reducing the 
classical dynamics to the fundamental domain which is the smallest part of the 
phase space which tesselates the whole space under application of the
allowed symmetry operations. After this procedure one obtains the
reduced density of states in the irreducible representation $m$
\beq \label{rde}
\delta g^m(E)=\frac{d_m}{\hbar}\sum_l \frac{\overline T_l}{|K_l|}
              \sum_r \frac{\chi_m(g_l^r)}{|\overline{\rm M}_l^r-
              \rm D_l|^{\frac12}} 
              \cos\left[\frac{r}{\hbar}\overline S_l(E)-
              \frac{\pi}{2}\overline\sigma_{rl}\right]\!.
\eeq
Here $d_m$ is the dimension and $\chi_m(g)$ the character of the symmetry 
operator $g$ in the irreducible representation $m$. The bars in Eq.\ \eq{rde} 
indicate that actions, periods, stability matrices and 
Maslov indices are calculated in the fundamental domain, while $g_l^r$ is the 
operator that relates the $r$-th repetition of the reduced orbit $l$ with its 
original lifted into the the whole phase space. $|K_l|$ is the order of the 
group $K_l$ which leaves every point of the orbit $l$ invariant. By the
definition of the fundamental domain, this is the identity for orbits that 
stay in the interior of the fundamental domain, while there can exist more 
than one operation for orbits that lie on the boundaries.
The matrix D$_l$ is block-diagonal in coordinates with blocks given by 
$d(gq)/dq$ with $g\in K$. This matrix is again the identity for interior 
orbits, but can be different for boundary orbits.

It is usually easier to solve the equations of motion in the whole space 
than in the fundamental domain, where one has hard-wall reflections. 
Given the classical quantities for the total space, the task is then 
to find their reduced counterparts (marked with bars in \eq{rde}).
Take a Hamiltonian of the form $H({\bf p},{\bf r})= {\bf
p}^2/2m+V({\bf r})$ which is invariant under
the point-group symmetry $G$. Suppose that the subgroup $H$ leaves the
$l$ orbit invariant (not pointwise), then the $l$ orbit can be divided 
into $|H|$ copies related by symmetry \cite{ref:thank1}. There will be 
$|G|/|H|$ copies of the orbit in the full phase space (if we consider 
time reversal, then there are $2|G|/|H|$ copies of orbits without
time-reversal symmetry). Therefore the lifted orbit should be equivalent 
to the $|H|=r$-th repetition of the reduced orbit (or to the $|H|/2$-th
repetition for time-asymmetric orbits, which become librating orbits 
in the fundamental domain, and the $|H|/|K|$-th repetition for
boundary orbits). Hence, all the classical quantities should be 
inter-related as
\beq 
S_l(E) = r \overline S_l(E)\,, \qquad  T_l(E)=r \overline{ T}_l(E)\,,  
           \qquad \sigma_l= r\overline\sigma_l\,, 
           \qquad {\rm M}_l=\overline{\rm M}_l^r\,,
\eeq
since they are invariant under point transformations. The only 
difficulty  remains to find out which of the roots of M$_l$ must be taken. 
E.g., for $|H|=2$ we have M$_l=\overline{\rm M}_l^2 $. Thus, if the 
eigenvalues of M$_l$ are ${\rm e}^{± u_l}$, those of $\overline{\rm  M}_l$ 
can be $±{\rm e}^{± u_l/2}$. 
On the other hand, we know that for two-dimensional Hamiltonian systems, 
hyperbolic orbits always have even Maslov indices, while elliptic and 
inverse-hyperbolic orbits always have odd Maslov indices \cite{ref:Sug}. 
We have observed that this rule can be reversed in the fundamental domain.
\begin{figure}[htbp]
\centerline{
\includegraphics[width=6cm]{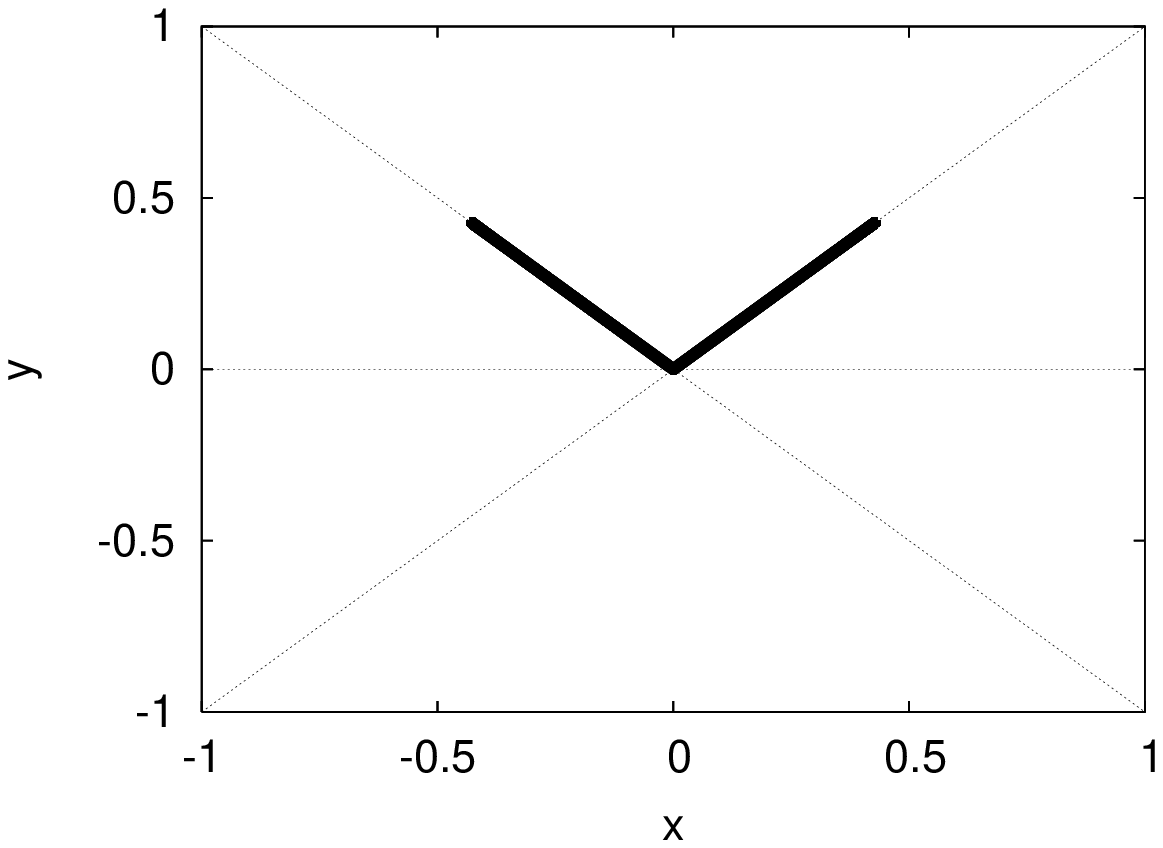}
\includegraphics[width=6cm]{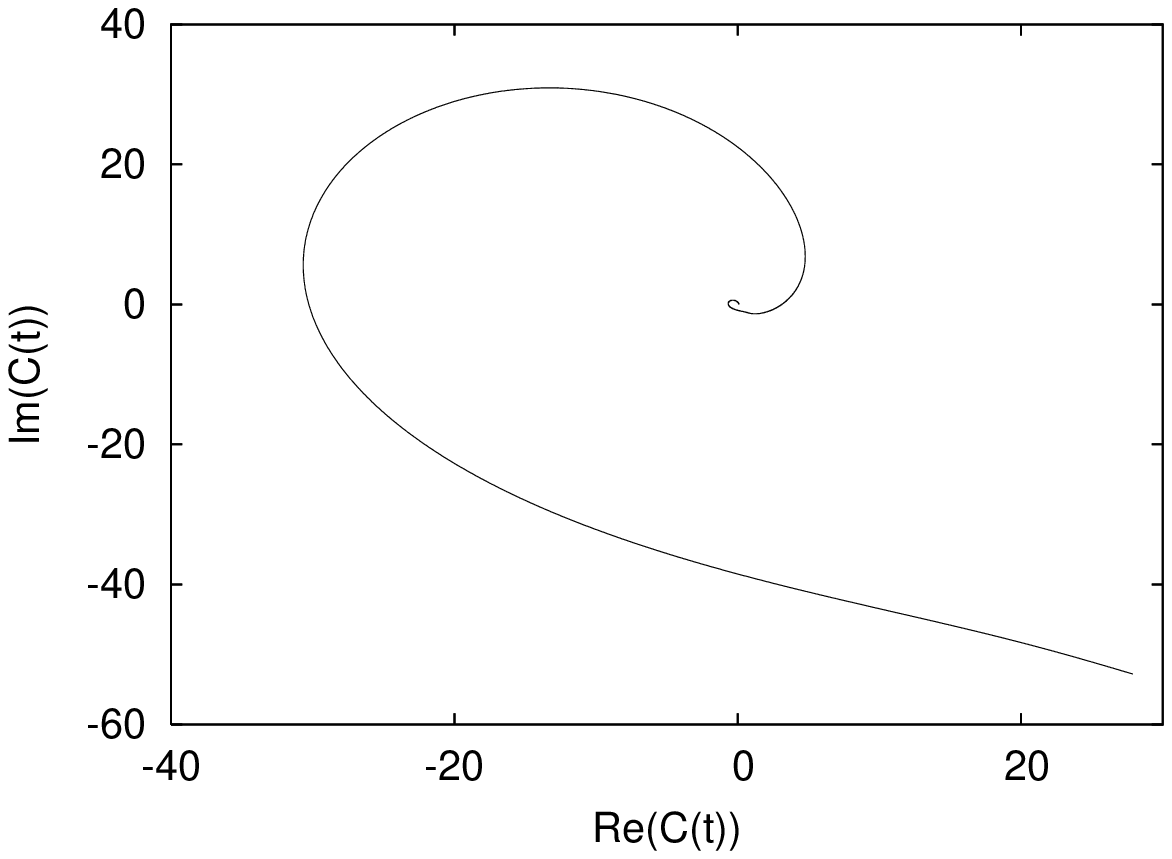}
}
\centerline{
\includegraphics[width=6cm]{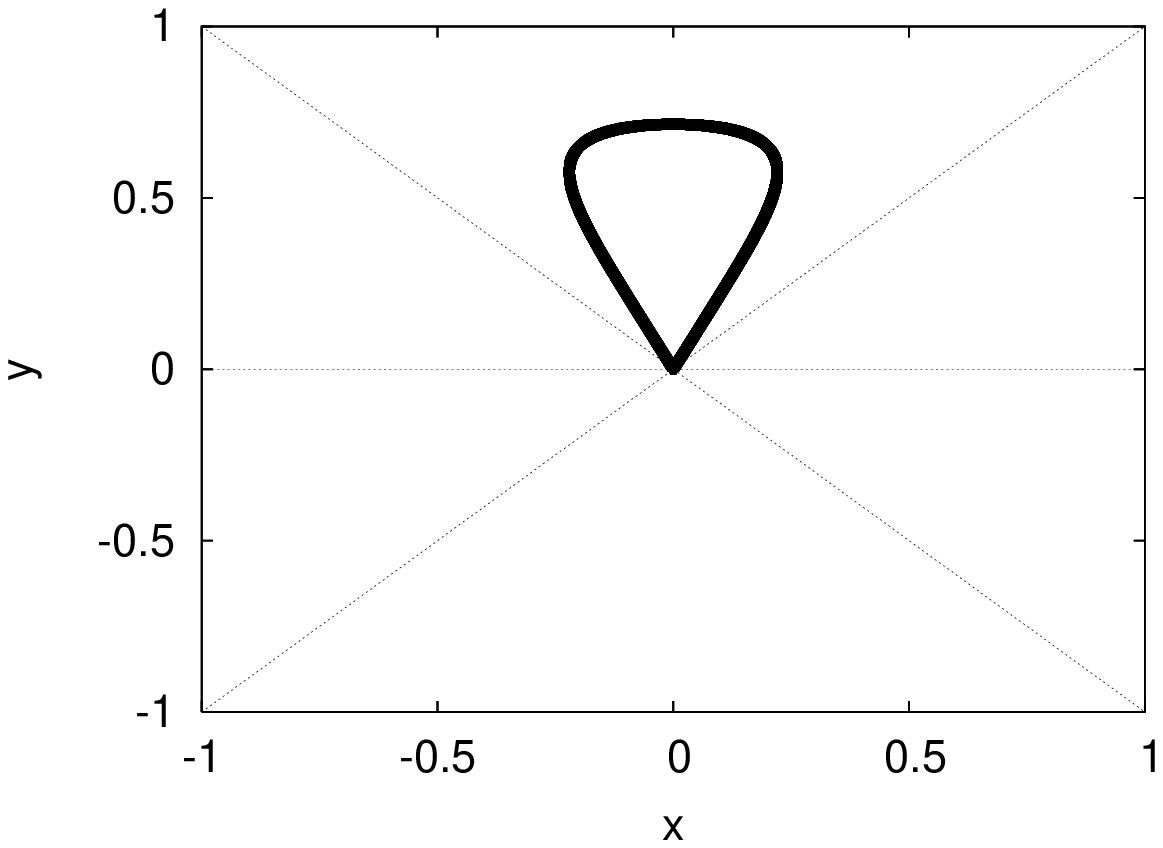}
\includegraphics[width=6cm]{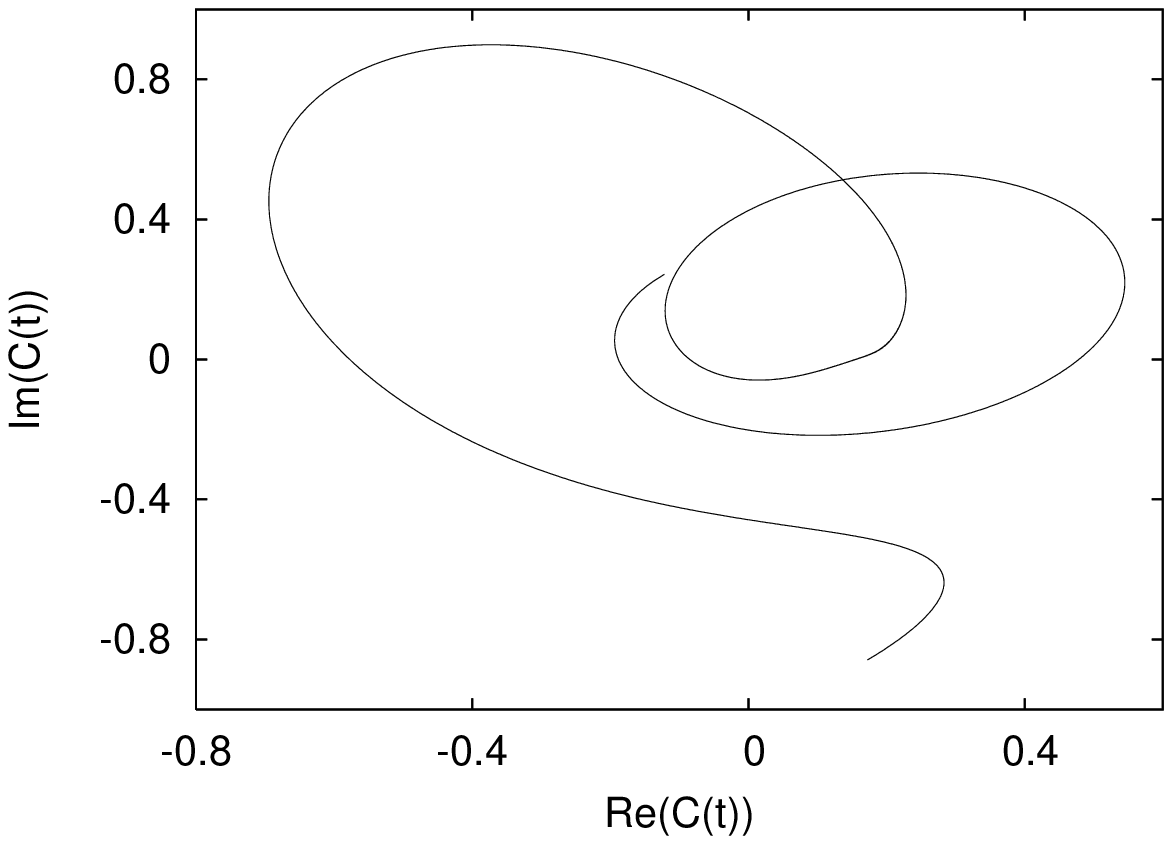}
}
\centerline{
\includegraphics[width=6cm]{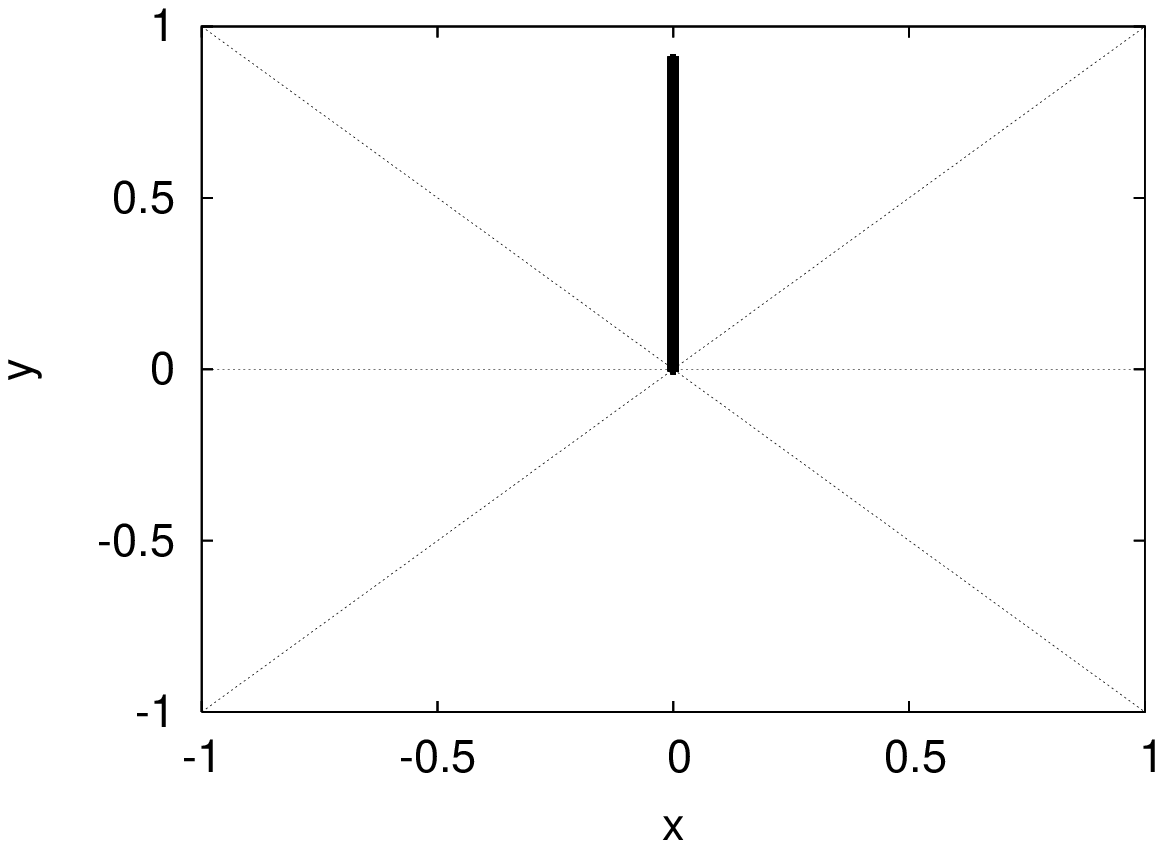}
\includegraphics[width=6cm]{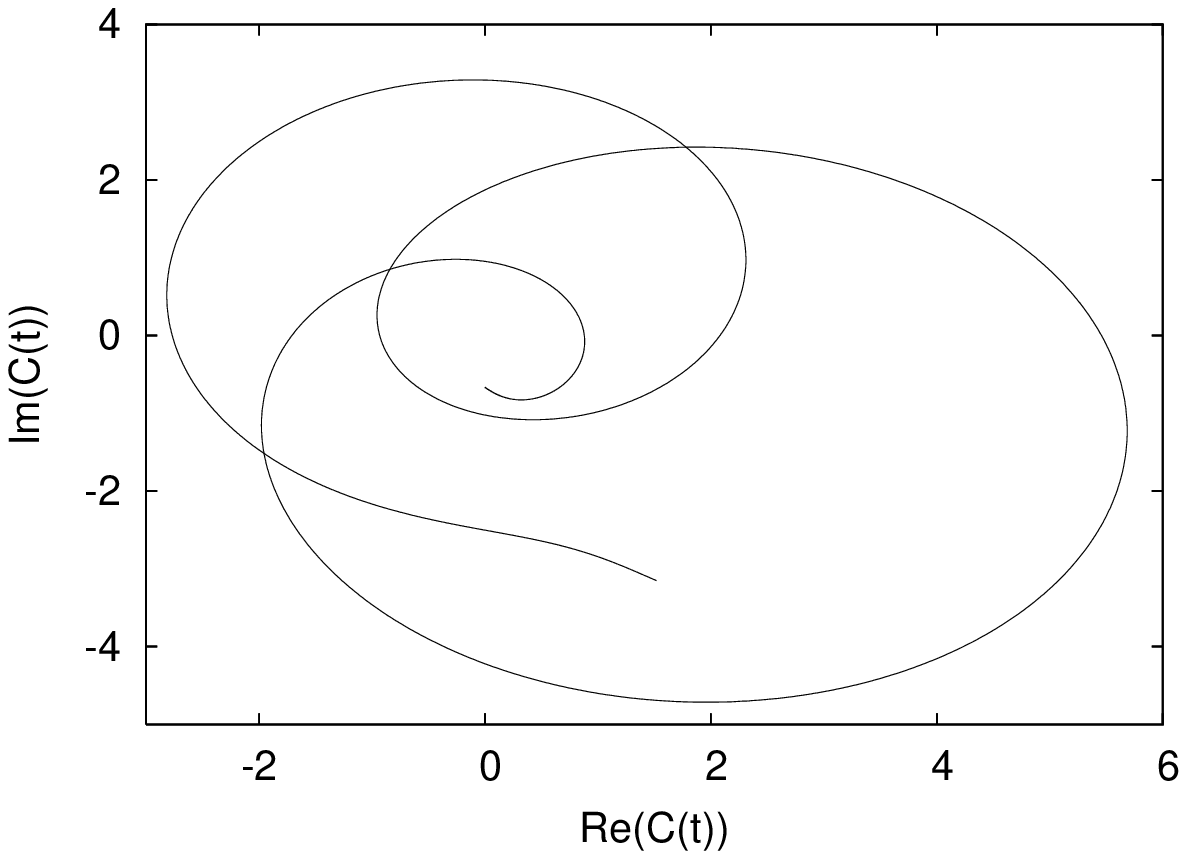}
}
\caption{Calculation of Maslov indices for some reduced orbits of the
QO at $\al=9$, considering only the reflexion symmetry at the $x$
axis. {\it Left panels:} reduced orbits in $(x,y)$ plane. 
{\it Right panels:} evaluation of the Maslov index $\overline\sigma$ 
which corresponds to the winding number of the complex number $C(t)$ 
over one period (cf.\ \cite{ref:crl}).
{\it Top panels:} librational orbit B$_2$ along the diagonal. Here the
length of the reduced orbit is the same as that of the lifted orbit,
and their Maslov indices are equal.
{\it Center panels:} orbit R$_4$. Here  the reduced orbit is half of 
the lifted orbit and its Maslov index is $\overline\sigma=2$ (i.e., half 
of the total $\sigma$) but Tr\,M is negative in spite of the even Maslov index.
{\it Bottom panels:} orbit $A_6$. The reduced orbit is again half of the 
total orbit, and so is the Maslov index. But Tr\,M is positive in 
spite of the odd Maslov index.}
\label{fig:mas}
\end{figure}

This is illustrated in \fig{fig:mas} for the case of a single reflection 
symmetry with respect to the $x$ axis. Then the fundamental domain 
is the upper plane ($y\geq0$). We have calculated the Maslov index 
$\overline\sigma$ using the method of Creagh {\it et al.} \cite{ref:crl} 
(as explained in Ref.\ \cite{ref:BraBa}, App.\ D) and verified that it is, 
indeed, either the same as for the lifted orbit for orbits without this 
symmetry, or half of it for orbits with reflection symmetry. However, 
the sign of the eigenvalues did not follow Sugita's rule \cite{ref:Sug}. 
This rule can, however, be applied to ${\overline \sigma}-mod(R,2)$, 
where $R$ indicates the number of hard-wall reflections at the
boundaries of the fundamental domain. Thus, if this number is odd, 
the rule is reversed. 

\begin{figure}[htbp]
\psfrag{E}{$E$}
\psfrag{dg_EES}{\large $\delta g^{EES}(E)$}
\begin{center}
\includegraphics[width=12cm]{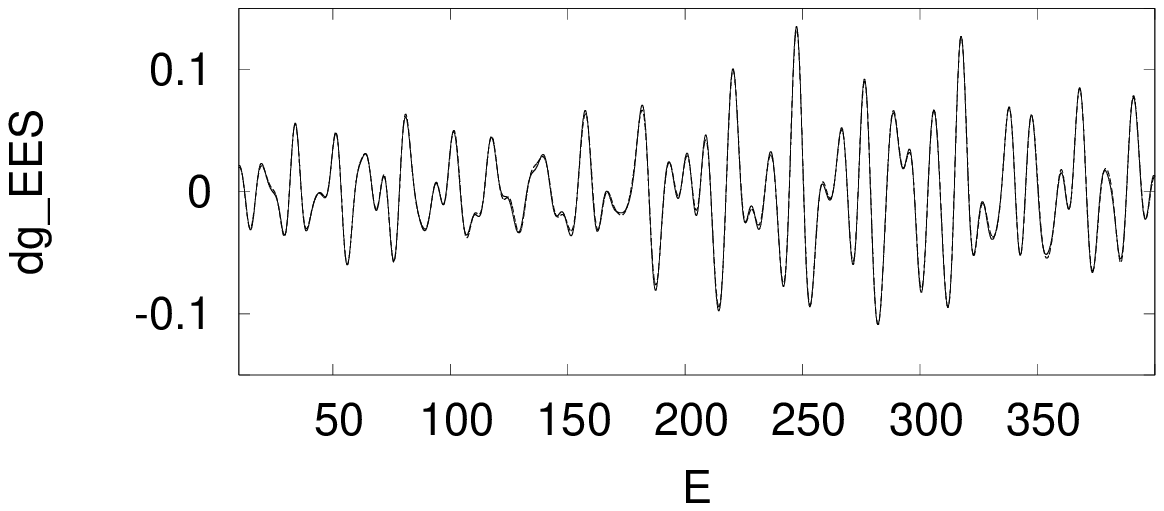}
\end{center}
\caption{Reduced density of states for the representation EES in 
the QO at $\alpha=9$ after Gaussian averaging with width $\gamma=4$.
The solid line shows the quantum result and the dotted line the
semiclassical result using Eq.\ \eq{rde}.}
\label{fig:rden9}
\end{figure}

We have calculated the reduced density of states \eq{rde} for the
representation EES in the QO at $\alpha=9$. The result is shown in 
\fig{fig:rden9} for Gaussian smoothing with width $\gamma=4$. A 
considerable agreement between the semiclassical (dotted line) and the 
quantum-mechanical result (solid line) is achieved.

\section{Spectral Rigidity}
\label{secrig}

To study the effect of pitchfork bifurcations on the spectral
statistics we consider the spectral rigidity or 
stiffness, $\Delta$ \cite{ref:DyMe}. It is defined as the local average of the 
mean-square deviation of the staircase function $N(E)$ from its best-fit
straight line over 
an energy range corresponding to $L$ states with mean level spacing $\bar{d}$:
\beq \label{delta3}
\Delta(L)=\Big\langle \min_{A,B}\, \frac{\bar{d}}{L}\!
          \int_{-L/2\bar{d}}^{L/2\bar{d}}{\rm d}\epsilon\, 
          [N(E+\epsilon)-A-B\epsilon]^2\,\Big\rangle.
\eeq
The quantity $\Delta (L)$  measures  spectral correlations over 
energy distances of order $L$. For an uncorrelated Poisson spectrum 
the universal prediction is 
\begin{equation}
\label{poisson}
\Delta^{\rm Poisson}(L)=L/15\,,
\end{equation}
while for a chaotic system it is approximately given by
\begin{equation}
\label{RMT}
\Delta^{\rm RMT}(L)=\frac{\beta}{2\pi^2}\log L-D\,,
\end{equation}
where D is a constant, $\beta=1$ for systems without time reversal symmetry (GUE
statistics) and $\beta=2$ for systems with time reversal symmetry (GOE statistics).
This universal behaviour has been observed up to correlation lengths 
$L<L_{\rm max}=2\pi\hbar \bar d/T_{\rm min}$, where $T_{\rm min}$ 
is the period of the shortest orbit.
\begin{figure}[htbp]
\psfrag{D(L)}{\small $\Delta(L)$}
\psfrag{delta(L)}{\small $\Delta(L)$}
\psfrag{Rigidity}{\small $\Delta(L)$}
\psfrag{L}{\small $L$}
\psfrag{N=10000}{\tiny\!\!\!\!\! $\tilde E=10000$}
\psfrag{N=8000}{\tiny\!\!\!\! $\tilde E=8000$}
\psfrag{N=6000}{\tiny \!\!\!\!$\tilde E=6000$}
\psfrag{N=4000}{\tiny \!\!\!\!$\tilde E=4000$}
\psfrag{N=2000}{\tiny \!\!\!\!$\tilde E=2000$}
\psfrag{N=1000}{\tiny\!\!\!\!\!\! $\tilde E=1000$}
\psfrag{N=500}{\tiny \!\!\!\!$\tilde E=5000$}
\begin{center}
\includegraphics[width=6cm]{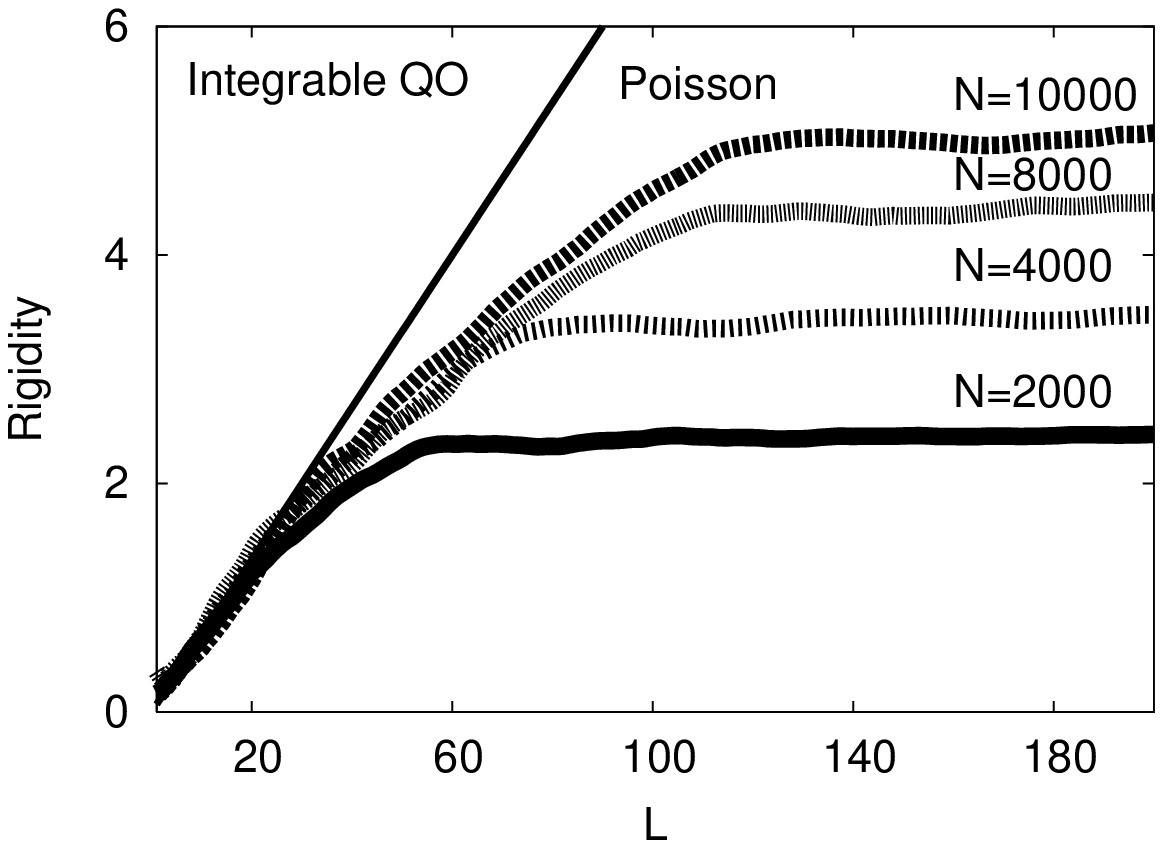}
\includegraphics[width=6cm]{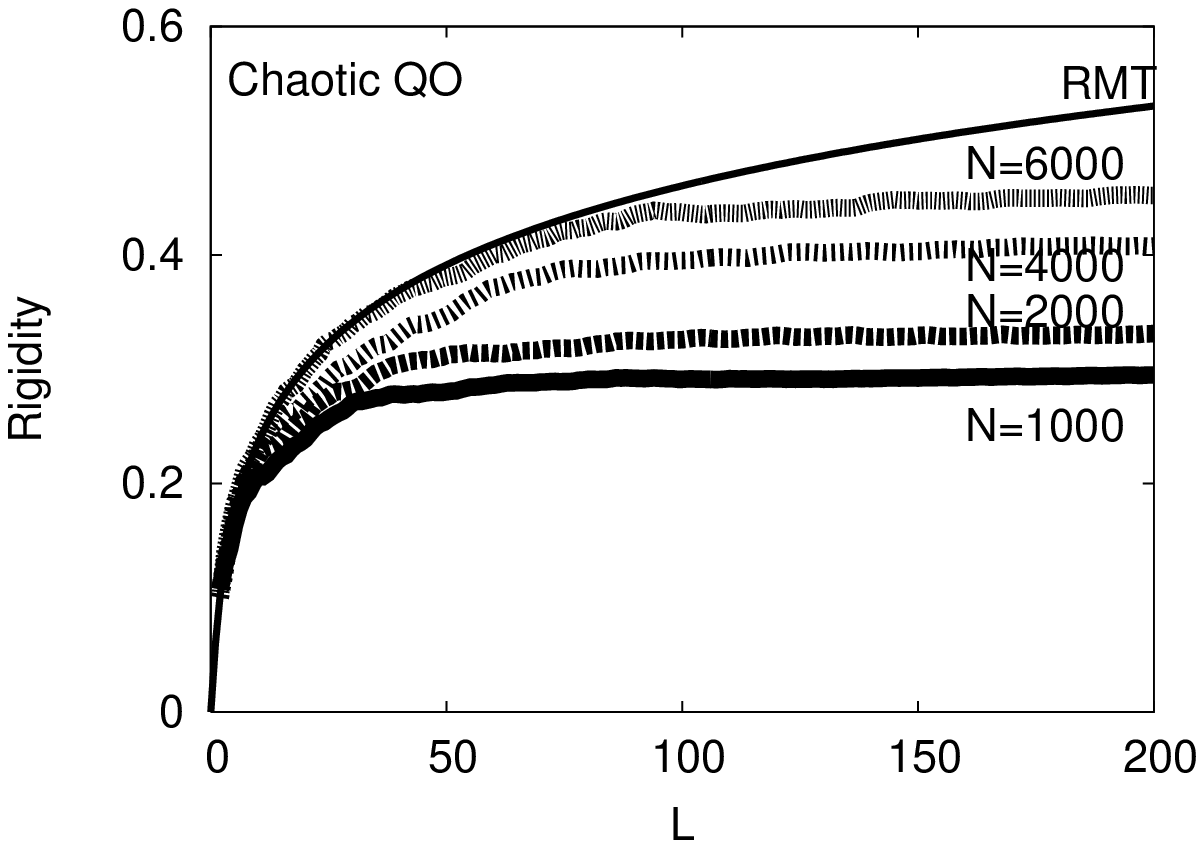}
\end{center}
\caption{Rigidity for $\al=0$ (integrable case) and $\al=9$ 
(almost chaotic case). With increasing (unfolded) energy
$\tilde E$  the numerical data converge to the 
universal Poisson (left panel) and random matrix predictions
(right panel) marked as full lines. }
\label{fig:rigex}
\end{figure}
In \fig{fig:rigex} we show the numerical results for the 
quartic oscillator in the integrable and almost chaotic
regime, compared with the corresponding  predictions, Eqs.\ 
(\ref{poisson}, \ref{RMT}). The $L$ range, in which the numerical 
data coincide with the universal predictions, increases with
increasing energy, i.e., by approaching the semiclassical limit.

For a mixed system it was conjectured that the statistics 
will be a superposition of Poisson and random matrix contributions 
\cite{ref:BerTa,ref:SelVer}, parameterized as


\beq \label{sumrule}
\Delta(L)\approx \Delta^{\rm Poisson}((1-q)L)+\Delta^{\rm RMT}(qL),
\eeq
were q is the irregularity fraction of the system (i.e., the fraction 
of the phase space corresponding to the chaotic sea). Since both 
statistics are monotonously increasing functions, we expect that the 
more regular the system is, the larger is the rigidity.

\subsection{Semiclassical theory for the rigidity}
\label{secsemrig}

\begin{figure}[htbp]
\psfrag{y}{\small $y$}
\psfrag{g(y)}{\small $g(y)$}
\centerline{
\includegraphics[width=6cm]{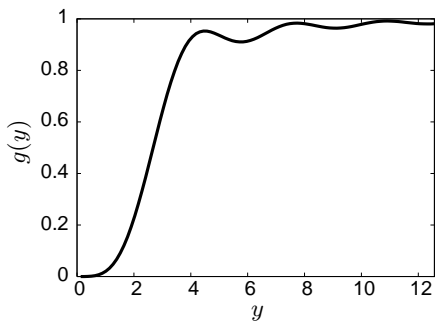}}
\caption{Window function $g(y)$, see text.
\label{fig:g}
}
\end{figure}

The semiclassical theory for the rigidity was developed in Ref.\
\cite{ref:Be85}, for the two 
limiting cases of complete chaoticity and full regularity (integrability).
The procedure is the following: By energy integration of the density 
of states, Eq.\ (\ref{dens}), one obtains an expression for the 
number of states. By inserting this expression into the
definition of the rigidity one finds
\beq \label{smr}
\Delta(L)=\frac{1}{2\hbar^{2\mu}} \Big\langle \!\sum_j\sum_k
  \frac{A_jA_k}{T_jT_k} \cos\!\left[\frac{1}{\hbar}(S_j-S_k)
  +\frac{\pi}{2}(\sigma_j-\sigma_k)\right]
  \!G(y_j,y_k)\!\Big\rangle,
\eeq
where $T_j=dS_j/dE$ are the periods, 
\beq y_j=\frac{LT_j}{2\hbar\bar d}=
         \pi\frac{L}{L_{\rm max}}\frac{T_j}{T_{\rm min}} \, ,
\eeq
and
\bea \label{G}
 G(x,y) & = & F(x-y)-F(x)F(y)-3F'(x)F'(y), \\
 F(x) & = & \frac{1}{x}\sin x = j_0(x)\,.
\eea
The main contributions come from pairs of orbits whose action 
difference is smaller than $\hbar$, so that $y_j$ can be chosen 
to be equal to $y_k$ in the argument of $G$:
\beq \label{smr2}
\Delta(L)=\frac{1}{2\hbar^{2\mu}}\Big \langle \sum_j\sum_k 
          \frac{A_jA_k}{T_jT_k}
          \exp\!\left[\frac{i}{\hbar}(S_j-S_k)+\frac{\pi}{2}(\sigma_j-\sigma_k)\right]
          g(\bar y_{j,k}) \Big\rangle,
\eeq
where $\bar y_{jk}=\frac{1}{2}(y_j+y_k)$ and $g(x)=G(x,x)$.
The function $g(y)$ (see \fig{fig:g}) selects the orbits that 
contribute to the double sum. If $L<<L_{\rm max}$ then $g(y)$ 
is almost unity only for long orbits, while for $L>L_{\rm max}$ 
the function is almost unity for all $y$, and the most important 
contributions to $\Delta(L)$ come from short orbits due to the factor $1/T^2$. 
Since we are interested in studying the effects of a bifurcation 
of one of the shortest orbits, we are going to concentrate on the 
saturation behaviour, which corresponds basically to the first 
moment of the staircase function. 

The rigidity can be written in terms of the spectral form factor $K(\tau)$
(the Fourier transform of the autocorrelation fuction) as
\beq\label{rigfofa}
\Delta(L)=\frac{1}{2\pi^2}\int \frac{K(\tau)}{\tau^2}g(\pi L\tau)d\tau, 
\eeq
with $\tau = T/2\pi\hbar\bar d$ and $K(\tau)=\left \langle\frac{1}{\bar d}
\int_{-\infty}^{\infty }\langle d(E+\omega/2)d(E-\omega/2)\rangle\,
e^{-2\pi i\omega\tau\bar d }{\rm d}\omega\right \rangle _{\Delta \tau}$. A local time average $\Delta \tau $ has to be performed in order to obtain a self-averaging form factor.

The corresponding semiclassical expression for the form factor, 
analogous to Eq.\ \eq{smr2}, is
\beq\label{smcfofa}
K(\tau,E)= \displaystyle \frac{1}{\hbar^{2\mu}}\Big\langle \!\sum_{j,k}\!
           \frac{A_jA_k}{T_H^2}\cos\!\left[\frac{1}{\hbar}(S_j-S_k)
  +\frac{\pi}{2}(\sigma_j-\sigma_k)\right]
   \!\delta_{\Delta \tau}\!\left(\!\tau-\frac{\bar T_{jk}}{T_H}\!\right)\!
   \Big\rangle_{\!\Delta E},
\eeq
where $\bar T_{jk}=\frac{1}{2}(T_j+T_k)$. The width of the delta-function is due to the time average $\Delta \tau $.

As expressed in Eqs.\ (\ref{smr2}, \ref{smcfofa}), 
the rigidity and the spectral form factor are determined by a double 
sum over pairs of periodic orbits. The semiclassical limit 
$\hbar\to 0$ means that the typical classical actions of these 
paths are very large compared with $\hbar$, so that the energy 
average will strongly suppress the contributions of most pairs 
of orbits. The first approximation is to consider that only 
orbits paired with themselves ($j=i$) or with their time-reserved 
partners ($j={\bar i}$) give a contribution, which is known as 
the ``diagonal approximation'' \cite{ref:Be85}.

For the QO at $\al=0$, the tori amplitudes $A_{kx,ky}$  are
given by
\beq
A_{kx,ky}=\left(\frac{{\bf K}}{\pi}\right)^{3/2}
          (4E)^{1/8}\frac{k_xk_y}{(k_x+k_y)^{5/8}}
\eeq
for the irreducible representation EES.
For integrable systems the contribution of the non-diagonal 
terms $j \neq k$ in the sum (\ref{smr}) will vanish after 
averaging owing to destructive interference. For this system, due to the degeneracy 
in the actions, the orbits that contribute to the 
double sum are those that satisfy $n_x^4+n_y^4=n_x'^4+n_y'^4$. 
Inserting the amplitudes for the tori and summing only over terms 
with the same actions we have
\beq \label{smrIQO}
\Delta(L)=\frac{(4E)^{3/4}}{2^4\pi^3\hbar{\bf K}} 
          \sum_{k_x,k_y=1}^\infty \!\!\frac{k_xk_y}{l_k^7}\,g(\bar y_{k_x,k_y})
          \!\!\sum_{n_x,n_y=1}^\infty \!\!n_xn_y\delta_{l_k-l_n}  \, ,
\eeq
where $l_k=(k_x^4+k_y^4)^{1/4}$, $l_n=(n_x^4+n_y^4)^{1/4}$, and
$\delta$ is the Kronecker delta.

\begin{figure}[htbp]
\psfrag{D(L)}{\small$\Delta(L)$}
\psfrag{L}{\small$L$}
\psfrag{Rigidity}{\small $\Delta(L)$}
\psfrag{N=10000}{\small \!\!\!\!\!\!\!\!\!\!$\tilde E=10000$}
\psfrag{N=8000}{\small \!\!\!\!\!\!\!$\tilde E=8000$}
\psfrag{N=6000}{\small \!\!\!\!\!\!\!$\tilde E=6000$}
\psfrag{N=4000}{\small \!\!\!\!\!\!\!$\tilde E=4000$}
\psfrag{N=2000}{\small \!\!\!\!\!\!\!$\tilde E=2000$}
\psfrag{N=1000}{\small \!\!\!\!\!\!\!$\tilde E=1000$}
\psfrag{N=500}{\small \!\!\!\!\!\!\!$\tilde E=500$}
\centerline{
\includegraphics[width=10cm]{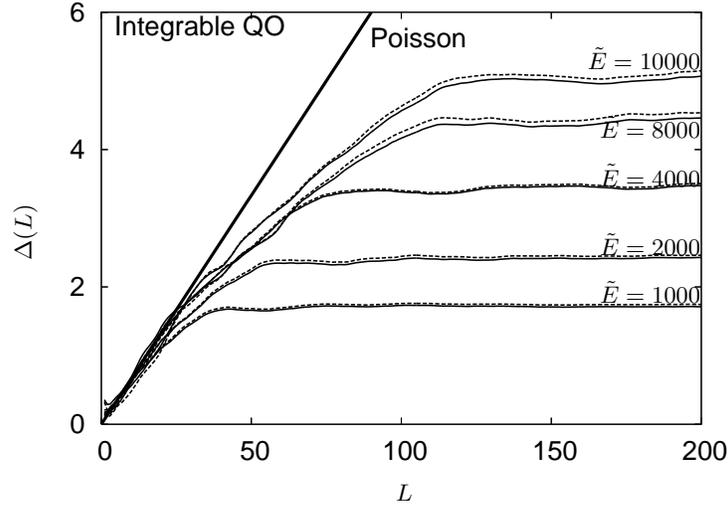}}
\caption{Semiclassical (dashed lines) and
quantum results (solid lines) for the spectral rigidity
$\Delta (L)$ for different values of $\tilde E$.
\label{fig:intrig}}
\end{figure}

With this expression, we can reproduce very well the statistics 
semiclassically, as is shown in \fig{fig:intrig}.

We have neglected here the contribution of the $A$ orbit which 
undergoes a pitchfork bifurcation. We have checked that its 
contribution is negligible, since its amplitude in the PO 
expansion goes like $\hbar^{-5/4}$ (a power one quarter larger than 
an isolated orbit) compared with that ($\hbar^{-3/2}$) of the tori.
For the saturation we can take $g(y_{k_x,k_y})=1$. Then the energy 
dependence of the saturation value $\Delta_\infty$ goes like 
$E^{3/4}$, as seen from Eq.\ \eq{smrIQO}. 

\begin{figure}[htbp]
\psfrag{Saturation}{\small $\Delta_\infty(\tilde E)$}
\psfrag{Sat(E)}{\small $\Delta_\infty(\tilde E)$}
\psfrag{alpha}{$\al$}
\psfrag{N}{\small $\tilde E$}
\psfrag{Lmax}{\small$L_{\infty}$}
\psfrag{Rigidity}{\small $\Delta(L)$}
\centerline{
\includegraphics[width=6cm]{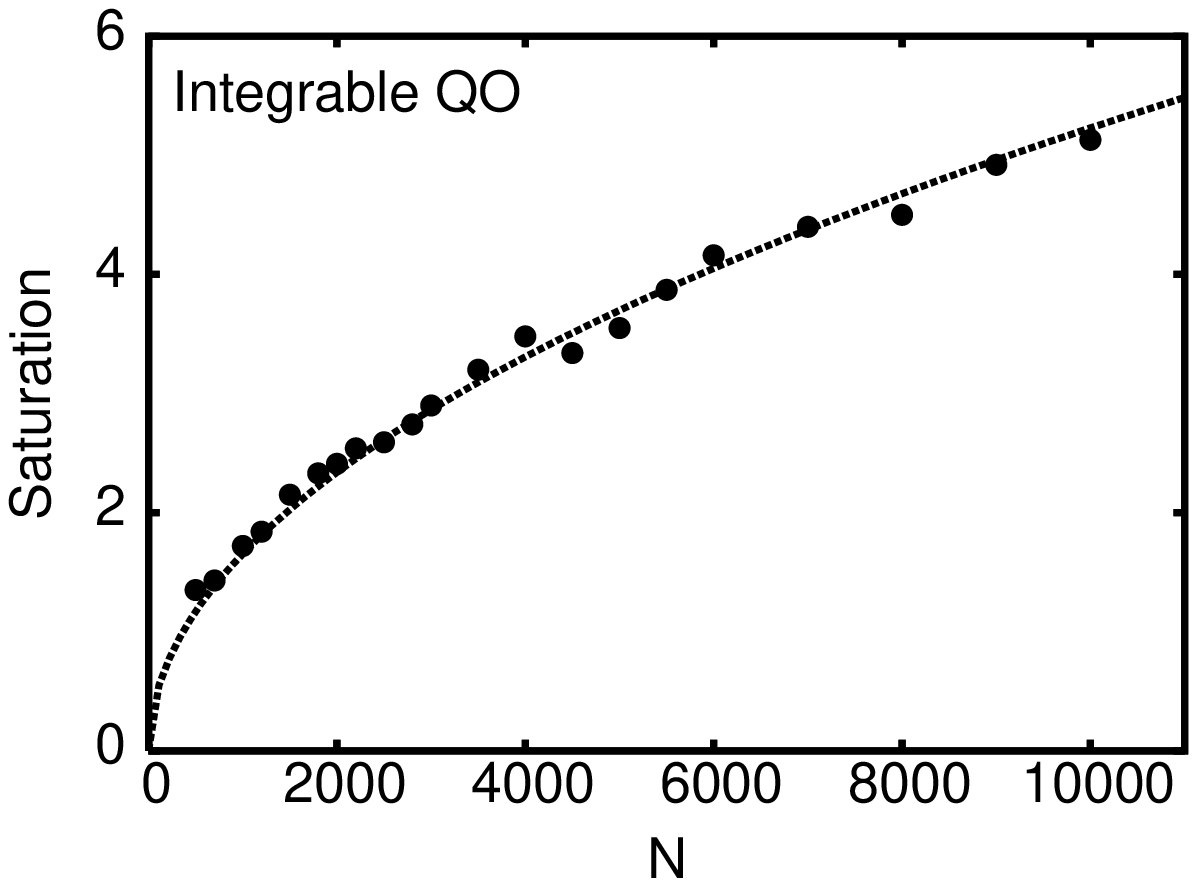}
\includegraphics[width=6cm]{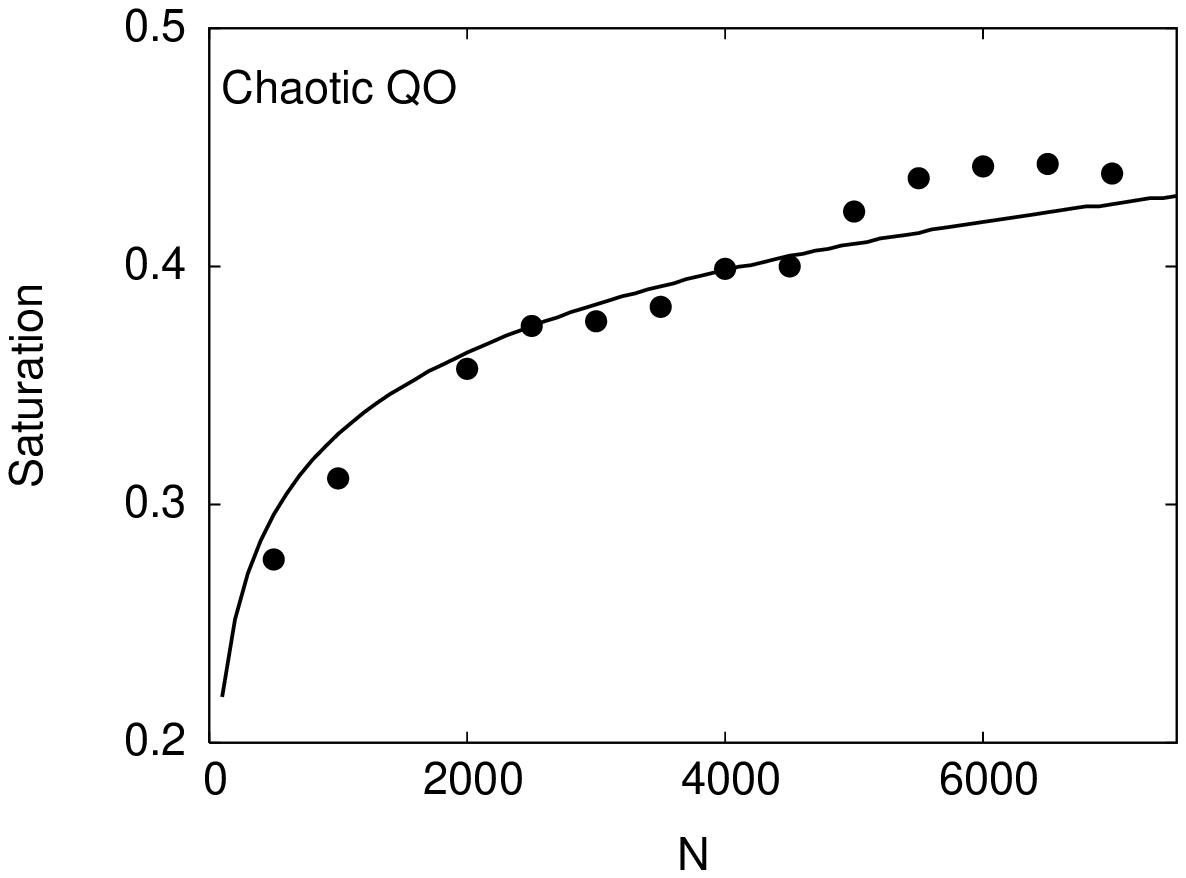}
}
\caption{Saturation value $\Delta_\infty$, plotted
versus unfolded energy $\tilde E$. The dots mark the quantum 
results. Left panel: integrable case $(\al=0)$; the solid line shows the 
semiclassical prediction. Right panel: Almost chaotic case ($\al=9$). 
Here the solid line represents the GOE prediction.}
\label{fig:int2}
\end{figure}

In the left panel of \fig{fig:int2} we depict the saturation value 
obtained from the quantum spectrum (dots), 
which is well reproduced by the semiclassical prediction (solid line).
For the chaotic case, RMT gives a saturation value $\Delta_\infty$
that behaves as $\log(1/\hbar)$ which is obtained by replacing the 
form factor by its GOE prediction in Eq.\ \eq{rigfofa}. Though
the exact saturation value is not universal, since it depends on the 
lower integration limit $\tau_{\rm min}$, its $\hbar$ dependence is.
In the right panel of \fig{fig:int2} we compare the quantum 
result with the GOE prediction evaluated for our value of $T_{\rm min}$ 
for $\al=9$.

\subsection{Bifurcation effects in the rigidity}

It has been discussed in Ref.\ \cite{ref:BeKeaPra} that 
additional contributions to the long-range spectral correlations 
may arise from bifurcations of periodic orbits, and that this 
effect can be reproduced semiclassically. The authors of Ref.\ 
\cite{ref:BeKeaPra} investigated the cat map at a 
tangent bifurcation, and found that the number variance of the 
counting function shows a  ``lift off'' 
reaching a much higher value than in the normal chaotic situation. 
We report here similar findings for the rigidity of the QO Hamiltonian for 
values of $\al$ near the pitchfork bifucations of the A orbit at
$\alpha_n$. Moreover, we find that the increase of the saturation 
value $\Delta_\infty$ becomes even larger slightly above the 
bifurcations.
\begin{figure}[htbp]
\psfrag{D(L)}{\small$\Delta(L)$}
\psfrag{delta(L)}{\small$\Delta(L)$}
\psfrag{L}{\small$L$}
\psfrag{Saturation}{\small $\Delta_\infty(\tilde E)$}
\psfrag{Sat(E)}{\small $\Delta_\infty(\tilde E)$}
\psfrag{N}{\small $\tilde E$}
\psfrag{alpha}{\small$\al$}
\psfrag{Rigidity}{\small $\Delta(L)$}
\psfrag{alpha=9}{\tiny $\al=9$}
\psfrag{alpha=10}{\tiny $\al=10$}
\psfrag{alpha=10.5}{\tiny $\al=10.5$}
\psfrag{alpha=11}{\tiny $\al=11$}
\centerline{
\includegraphics[width=6cm]{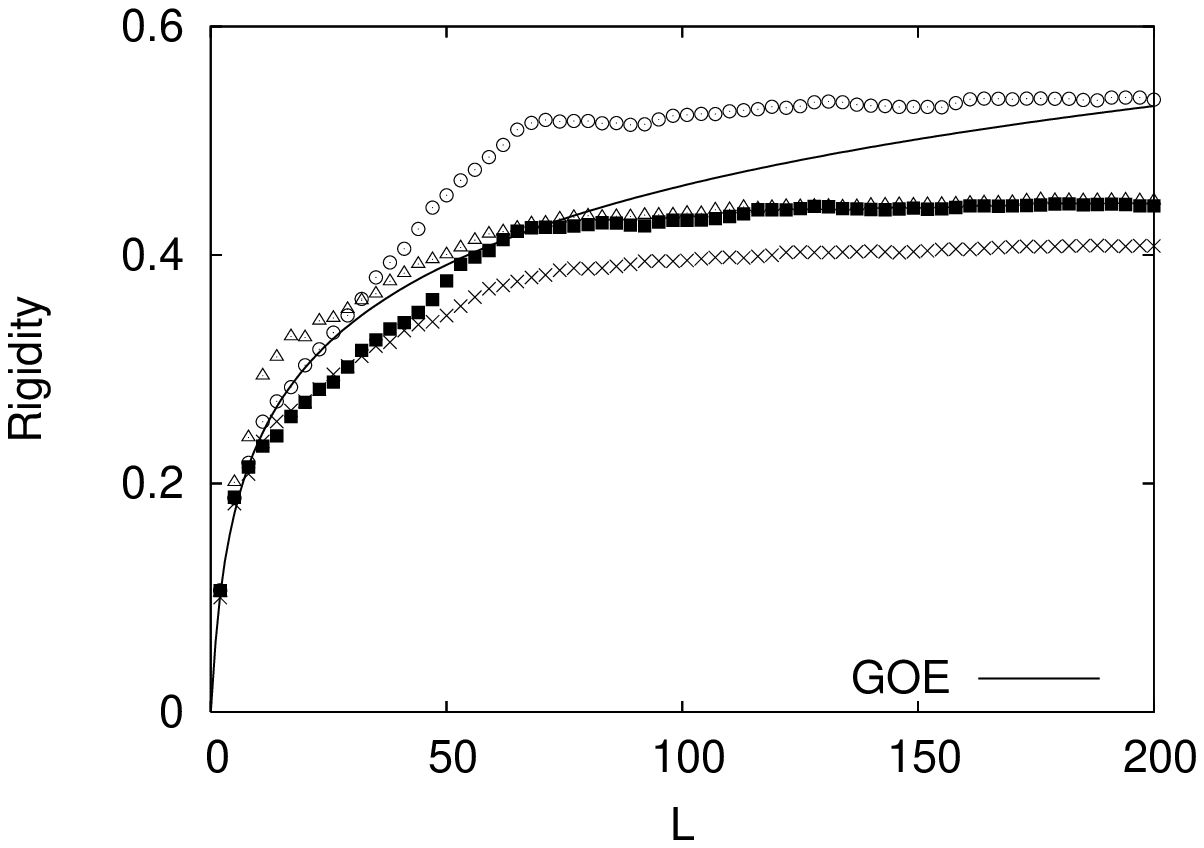}
\includegraphics[width=6cm]{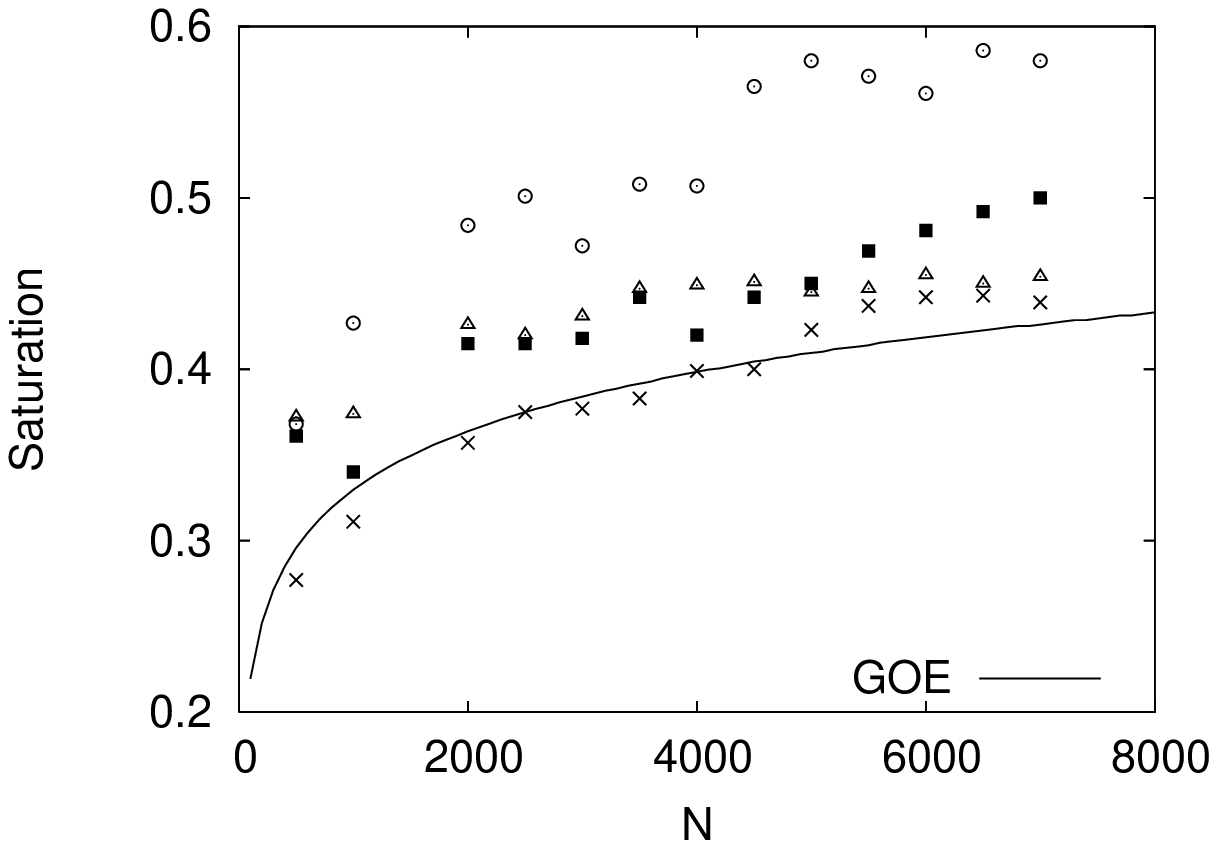}
}
\caption{Left panel: spectral rigidity for $\al=9$ (crosses), 
$\al=\al_4=10$ (filled squares), $\al=10.5$ (circles), and $\al=11$ 
(triangles) for $\tilde E=4000$. Right panel: saturation value $\Delta_\infty$
versus $\tilde E$ before and after the bifurcation at $\alpha_4 = 10$. 
Although the phase space is barely affected, the saturation at 
$\al=10.5$ is much larger than the saturation at $\al=11$.
\label{fig:dbif}}
\end{figure}
\begin{figure}[htbp]
\psfrag{K(t)}{\small $K(\tau)$}
\psfrag{t}{\small $\tau$}
\centerline{
\includegraphics[width=6cm]{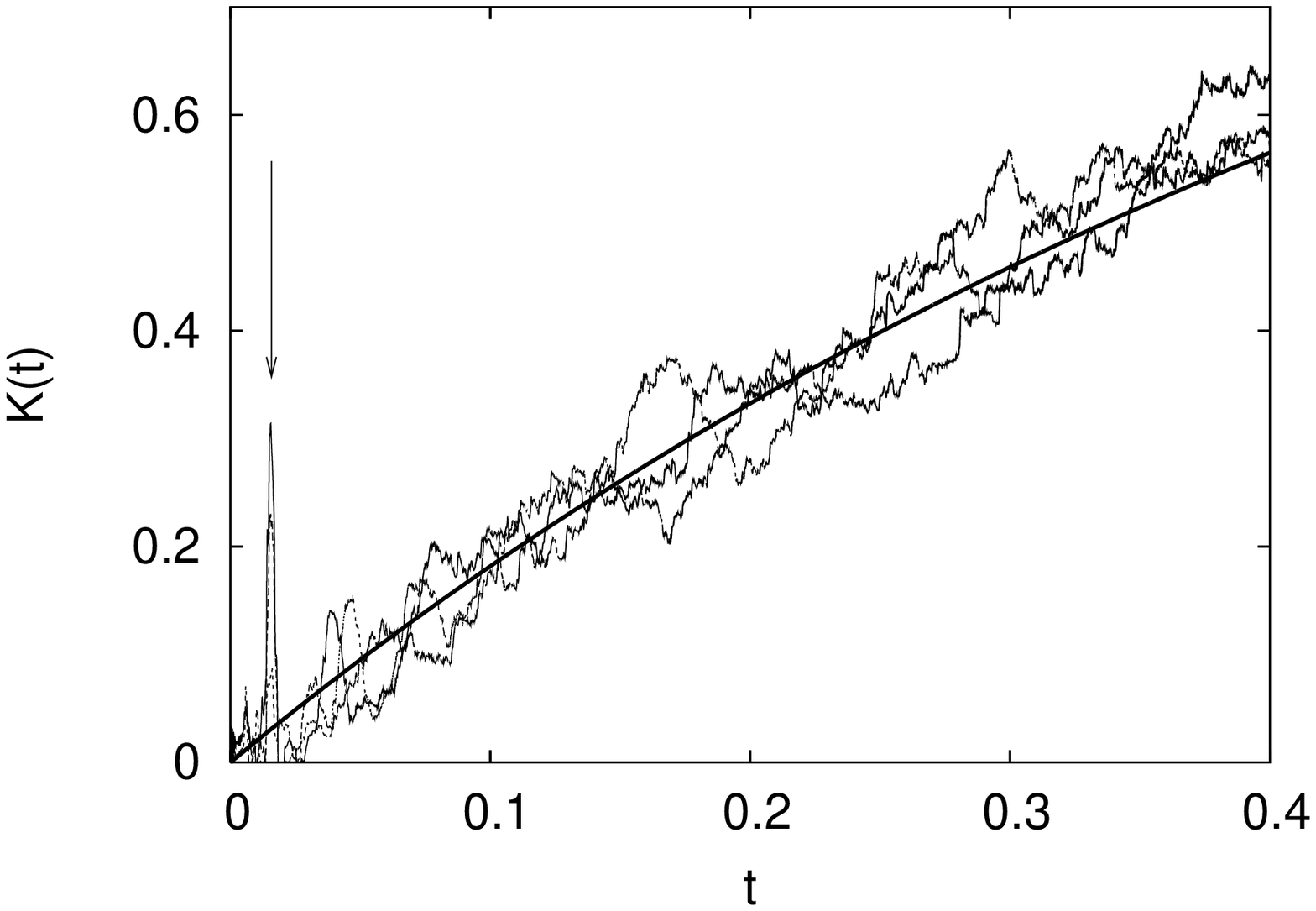}
\includegraphics[width=6cm]{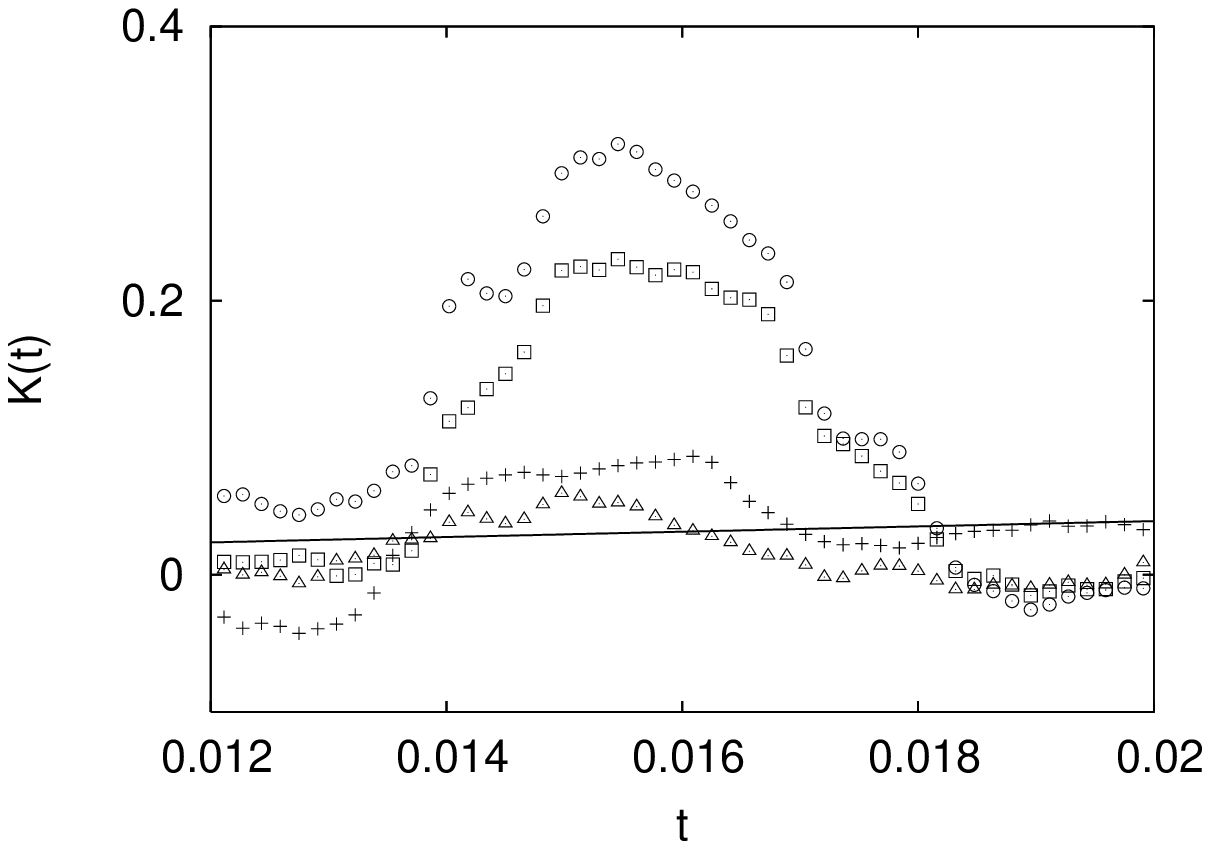}}
\caption{
{\it Left panel:} Form factor at $\al=9, 10$ and $10.5$ compared 
with RMT (line). Note the strong peak at $\tau=\tau_{\rm A}\simeq
0.015$ (indicated by an arrow) coming from the bifurcating orbit. 
{\it Right panel:} Form factor at $\al=9$ (crosses), $\al=10$ (squares), 
$\al=10.5$ (circles), and $\al=11$ (triangles) in a zoomed region
around $\tau\simeq\tau_{\rm A}$. For $\al=10.5$: the amplitude 
of the peak is clearly larger than at the bifurcation. 
\label{fig:dbiffofa}}
\end{figure}
This is illustrated in \fig{fig:dbif}. In the left panel we 
show the rigidity $\Delta(L)$ for four values of $\al$ around
$\al=\al_4=10$ where such a bifurcation occurs. The 
rigidity at $\al_4=10$ exhibits a slightly larger saturation than 
at $\al=9$ (``lift off''). However, the increase is even
much more noticeable at $\al=10.5$. Then the saturation goes 
down again for $\al=11$, even though the system is more 
regular than at $\al=10.5$ \cite{ref:nota1}.

\begin{figure}
\psfrag{px}{\small $p_x$}
\psfrag{x}{\small $x$}
\psfrag{alpha}{$\al$}
\psfrag{a=9}{\tiny $\al=9$}
\psfrag{a=10}{\tiny $\al=10$}
\psfrag{a=10.5}{\tiny $\al=10.5$}
\psfrag{a=11}{\tiny $\al=11$}
\psfrag{N}{\tiny $\tilde E$}
\centerline{
\includegraphics[width=12cm]{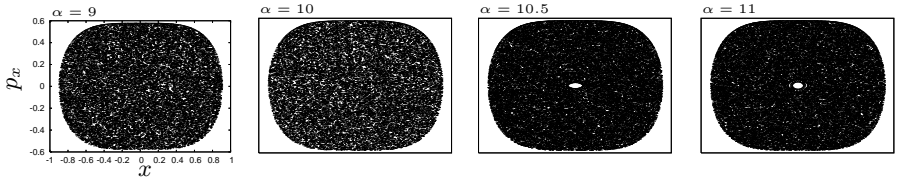}
}
\caption{Poincaré surfaces of section for the QO near $\al_4$. 
At $\al=9$ and at the bifurcation point $\al=10$ the phase space 
looks competely chaotic. 
A new stable island appears at $\al=10.5$, which is slightly 
larger at $\al=11$. 
}
\label{fig:pssbif}
\end{figure}

The energy dependence of $\Delta_\infty$ is shown in the right 
panel of \fig{fig:dbif}. We see that this effect exists over a large 
region of energies. As depicted in \fig{fig:pssbif} the phase space 
looks completely chaotic at the 
bifurcation at $\al=10$; without knowledge of the 
bifurcation one would expect an almost universal behaviour. Above the 
bifurcation, a tiny regular island is seen at the center, which
arises from orbit A$_7$ that  became stable. The 
island is slightly larger at $\al=11$ than at $\al=10.5$ (see 
\fig{fig:pssbif}).

Equivalently, in \fig{fig:dbiffofa} we show the effect in the 
spectral form factor. In the left panel we show $K(\tau)$ at 
$\al=9, 10$ and $11$. The results are consistent with the GOE prediction for 
almost all times, but we see a very large peak at a time that 
corresponds to the period of the libration orbit, $\tau_{\rm A}$. 
This is consistent with the results of \cite{ref:BeKeaPra}. 
However, the enhancement is even more noticeable at $\al=10.5$ 
(right panel).

The exact calculation of the semiclassical rigidity for the
QO in the chaotic regime is numerically impossible, since this
would require an infinite number of periodic orbits, 
and there is no analytical way to calculate them. 
To reproduce the quantum result semiclassically, 
we calculate the coarse-grained reduced density of states,
defined analogously to Eq.\ (\ref{smcg}) by
\beq \label{rdecon}
\delta g^m_\gamma(E)=\frac{d_m}{\hbar}\sum_l 
              \frac{\overline T_l}{|K_l|}e^{-(\gamma \overline T_l/2)^2}
              \sum_r \frac{\chi_m(g_l^r)}{|\overline{\rm M}_l^r-
              \rm D_l|^{\frac12}} 
              \cos\left[\frac{r}{\hbar}\overline S_l(E)-
              \frac{\pi}{2}\overline\sigma_{rl}\right]\!.
\eeq
The longer orbits will be exponentially suppressed assuring
convergence, but, at the same time, affecting the universality. 
However, for the study of the saturation
properties of $\Delta(L)$ as a probe for bifurcation effects,
the information of the shorter orbits should be sufficient.

\begin{figure}[htbp]
\psfrag{Rigidity}{\small $\Delta_{\gamma}(L)$}
\psfrag{sat}{\small $\Delta_{\gamma}(L)$}
\psfrag{Saturation}{\small $(\Delta_\gamma)_\infty(\tilde E)$}
\psfrag{alpha}{$\al$}
\psfrag{N}{\small $\tilde E$}
\psfrag{L}{\small$L$}
\psfrag{alpha=9}{\tiny $\al=9$}
\psfrag{alpha=10}{\tiny $\al=10$}
\psfrag{alpha=10.5}{\tiny $\al=10.5$}
\psfrag{alpha=11}{\tiny $\al=11$}
\psfrag{gamma=1}{\tiny $\gamma=1$}
\psfrag{gamma=4}{\tiny $\gamma=4$}
 \centerline{
\includegraphics[width=6.0cm]{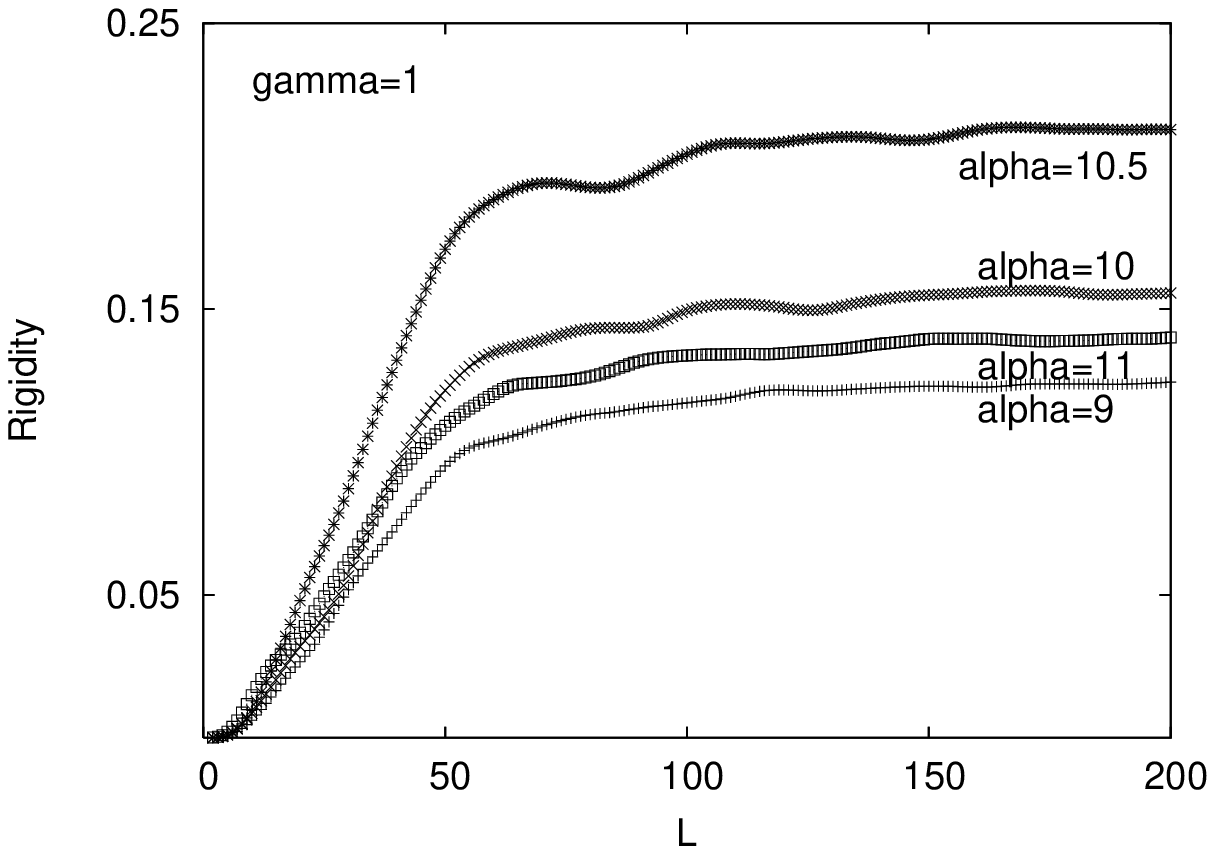}
\includegraphics[width=6.0cm]{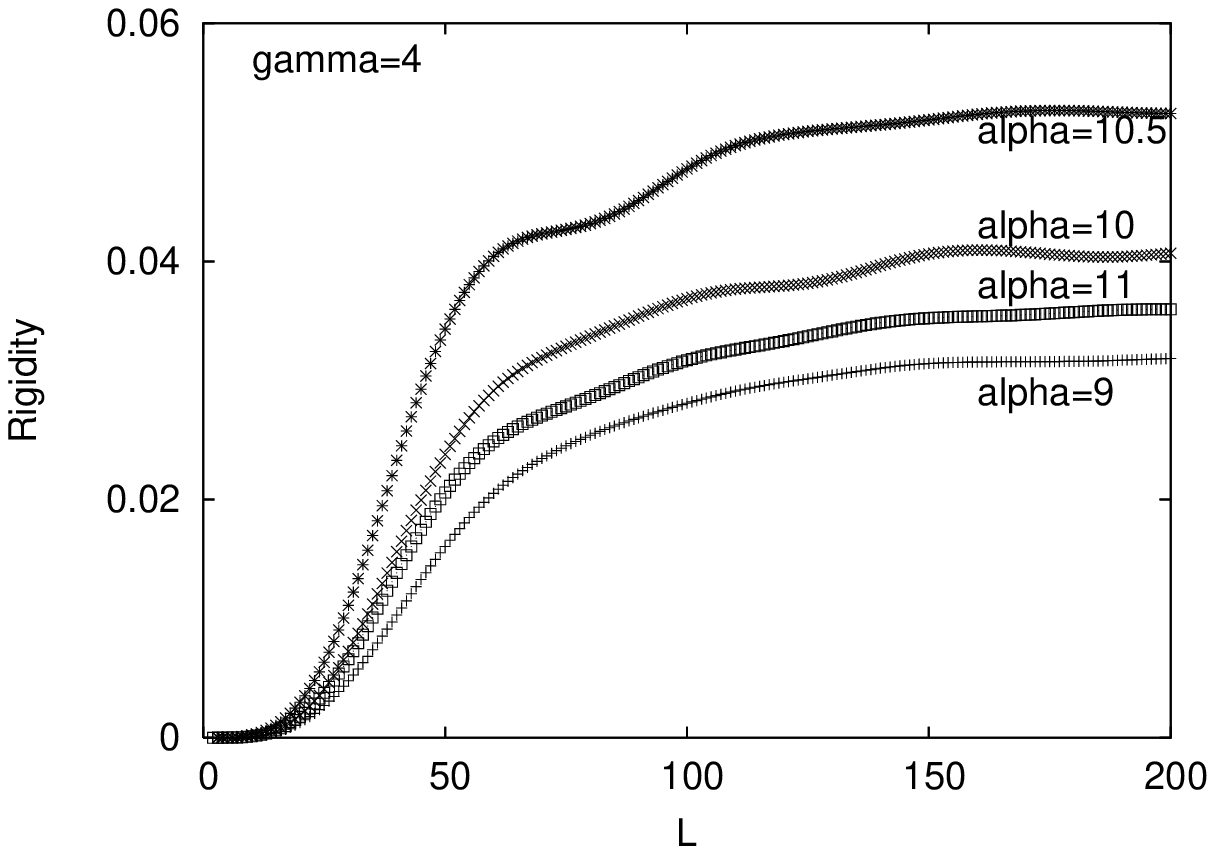}
}
\caption{Same as \fig{fig:dbif}, but after coarse-graining the
reduced quantum spectrum by a Gaussian  smoothing with width
$\gamma=1$ (left) and $\gamma=4$ (right).}
\label{fig:satsmoo}
\end{figure}

Consistently we also coarse-grain the quantum stair-case function, defining
\beq\label{stairg}
N_\gamma(E)=\frac12\sum_n\left[1-{\rm erf}\left(\frac{E_n-E}{\gamma}\right)\right]\,.
\eeq
Inserting  $N_\gamma(E)$ into Eq.\ \eq{delta3}, we obtain a ``smoothed'' 
rigidity $\Delta_\gamma$ of the coarse-grained density of states.
We find that even for relatively large values of $\gamma$, the 
bifurcation effects described above are still clearly visible, as
shown in \fig{fig:satsmoo}.

We are now equipped to calculate the saturation property of the
smooth rigidity $\Delta_\gamma(L)$ semiclassically, taking into account
the bifurcation of the A orbit at $\al_4=10$. Its contribution
to the total density of states,
together with that of the L$_6$ orbits born at the bifurcation,
to the total density of states is given in the ``global'' 
uniform approximation of Ref.\ \cite{ref:SchoSie} (with $\sigma=+1$,
$a<0$, $\sigma_1=-1$ and $\nu=\sigma_{\!A_6,r}=6r$ for the present 
case). It reads
\bea\label{gloapp}
\delta g^{un}_{A+L}(E) &\!\! = \!\!& \Re e \,\frac{1}{\pi\hbar}\left| 
   \frac{\pi \Delta S}{2\hbar} \right|^{1/2}
   \exp\left(\frac{i}{\hbar}\overline{S}-i3r\pi
             -i\frac{\pi}{4}\right)×\nonumber\\
& & × \left\{ {\overline A}
            \left[\sigma_2J_{1/4}\left(\!\frac{|\Delta S|}{\hbar}\!\right)
                        e^{-i\frac{\pi}{8}}
                 +J_{-1/4}\left(\!\frac{|\Delta S|}{\hbar}\!\right)
                        e^{i\frac{\pi}{8}}\right] +
    \right.\nonumber\\
& & \left. + \Delta A\!
             \left[J_{3/4}\left(\!\frac{|\Delta S|}{\hbar}\!\right)
                        e^{-3i\frac{\pi}{8}}
                 +\sigma_2J_{-3/4}\left(\frac{|\Delta S|}{\hbar}\right)
                        e^{3i\frac{\pi}{8}}\right]\!
    \right\}\!.
\eea
Here $\Delta A=A_L/2-A_A/\sqrt{2}$, ${\overline A}=A_L/2+A_A/\sqrt{2}$,
${\overline S}=(S_L+S_A)/2$ and $\Delta S=(S_L-S_A)/2$, where
$A_j(E)$ and $S_j(E)$ are the Gutzwiller amplitudes and actions of the
isolated A and L orbits, respectively, away from the bifurcation, $r$
is their repetition number, and $\sigma_2={\rm sign}(\al-\al_4)$. 
At the bifurcation ($\al=\al_4=10$), the local uniform approximation 
becomes
\beq\label{locun}
\delta g^{loc}_{A+L}(E)\,=\,\frac{T_A\Gamma(\frac14)}
           {2\pi\sqrt{2\pi}\,\hbar^{5/4}|a|^{1/4}r^{3/4}}
           \cos\left[\frac{S_A}{\hbar}-3r\pi
           +\frac{\pi}{8}\right]\!.
\eeq
Here $T_A(E)$ is the period of the primitive A orbit, and $a$ is 
a normal form parameter which we determined numerically
from the local expansion given in Eq.\ \eq{locact} below (cf.\ also
Ref.~\cite{ref:BraFeMagMeh,ref:SchoSie}).
 
\begin{figure}[htbp]
\psfrag{Delta(L)}{\small $\Delta_{\gamma}(L)$}
\psfrag{Rigidity}{\small $\Delta_{\gamma}(L)$}
\psfrag{L}{\small$L$}
 \centerline{
\includegraphics[width=8cm]{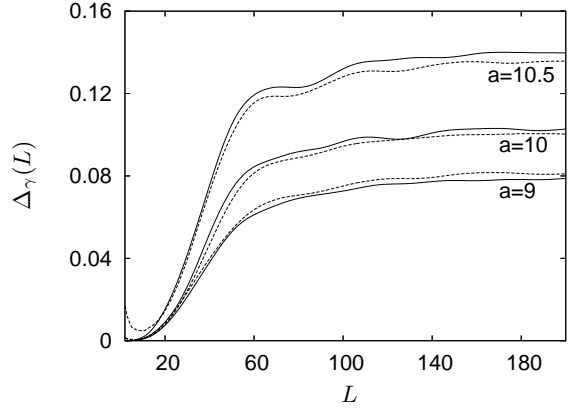}
}
\caption{Smoothed rigidity for $\al=9$, $\al=\al_4=10$ and $\al=10.5$, 
obtained for $\gamma=2$. The solid (dashed) curves represent the
quantum mechanical  (semiclassical) results. 
\label{fig:rigbif}}
\end{figure}

In this way we can reproduce the quantum mechanical results 
near the bifurcation semiclassically, as demonstrated in \fig{fig:rigbif}. 
Further analysis showed that amplitudes and actions of most of
the orbits do barely change, and the higher 
saturation for the smooth rigidity was mainly caused by the bifurcation. 

Considering the rigidity without smoothing, we now assume that
the contribution of the long orbits corresponds to and can be replaced by
the universal RMT prediction, so that the differences in the saturations 
arise basically from the A and L orbits.
Hence, we approximate the saturation value of $\Delta$ by 
\bea
\Delta_\infty(E) & \simeq & \Delta_{\infty}^{GOE}+\Delta_{\infty}^{A,L}\nonumber \\
&\simeq &\Delta^{GOE}+\frac12\,\Big\langle\!\!\sum_{j,k= A,L}\frac{A_{j}A_{k}}{T_{j}T_{k}}
                      \cos\left(\frac{S_j-S_k}{\hbar}\right)\!\Big\rangle \, .
\eea
At the bifurcation, the second term corresponds 
to the diagonal contribution of \eq{locun}, so that
\beq\label{sat10a}
\Delta_{\infty}^{A,L}=\frac{\Gamma^2(1/4)}{8\pi^3|a|^{1/2}\hbar^{1/2}} \, ,
\eeq
and $\Delta_\infty$ behaves like
\beq\label{sat10b}
\Delta_\infty\;\propto\; \log(1/\hbar)+\frac{1}{\hbar^{1/2}}.
\eeq
In the neighborhood of the bifurcation, i.e., when the action difference 
$|\Delta S|$ is smaller than $\hbar$, we can expand the actions 
and amplitudes around $\alpha=\alpha_4$ (cf.\ Ref.\ \cite{ref:SchoSie}):
\beq\label{locact}
\Delta S = \frac{S_{A}-S_{L}}{2} = \frac{\epsilon^2}{4a} + O(\epsilon^3)\,,
\eeq
\beq
A_{A} = \frac{T_{A}}{\sqrt{2\epsilon}}\,,\qquad 
A_{L} = \frac{T_{A}}{\sqrt{\epsilon}}\left[1+O(\epsilon) \right], 
\eeq
where $\epsilon=c\,(\alpha-\alpha_4)$. Up to first order in 
$\epsilon$ this yields
\bea\label{locex}
\delta g^{un}_{A+L}(E) & \approx & \frac{T_{A}}{\pi\sqrt{2\pi}\hbar}\,\Re e\, 
                             e^{i\bar S/\hbar -i3k\pi-i\pi/4}\,× \nonumber\\
       & &  ×   \left[\frac{\sigma_2\Gamma(3/4)}{|a\hbar|^{3/4}}\epsilon
          \,{\rm e}^{-i\pi/8}+\frac{\Gamma(1/4)}{2|a\hbar|^{1/4}}\,
          {\rm e}^{i\pi/8}\right]\!.
\eea
Inserting this into the saturation value of the rigidity we obtain
\beq\label{sat105}
\Delta_{\infty}^{A,L} \approx \frac{\Gamma^2(1/4)}{8\pi^3|a|^{1/2}\hbar^{1/2}}
                        +  \epsilon\,\frac{1}{2\pi^2|a|\hbar}
                        +  \epsilon^2\,\frac{\Gamma^2(3/4)}{2\pi^3|a|^{3/2}\hbar^{3/2}}\,.
\eeq
Equivalent results are obtained for the form factor considering only
the contributions of the orbits involved in the bifurcation.

\begin{figure}[htbp]
\psfrag{Saturation}{\small $\Delta_\infty(\tilde E)$}
\psfrag{Delta_infty}{\small $\Delta_\infty(\tilde E)$}
\psfrag{alpha}{$\al$}
\psfrag{N}{\small $\tilde E$}
\psfrag{Lmax}{\small$L_{\infty}$}
\psfrag{a=9}{\tiny $\al=9$}
\psfrag{a=10}{\tiny $\al=10$}
\psfrag{a=10.5}{\tiny $\al=10.5$}
\psfrag{y}{\small $K(\tau)$}
\psfrag{t}{\small $\tau$}
 \centerline{
\includegraphics[width=6cm]{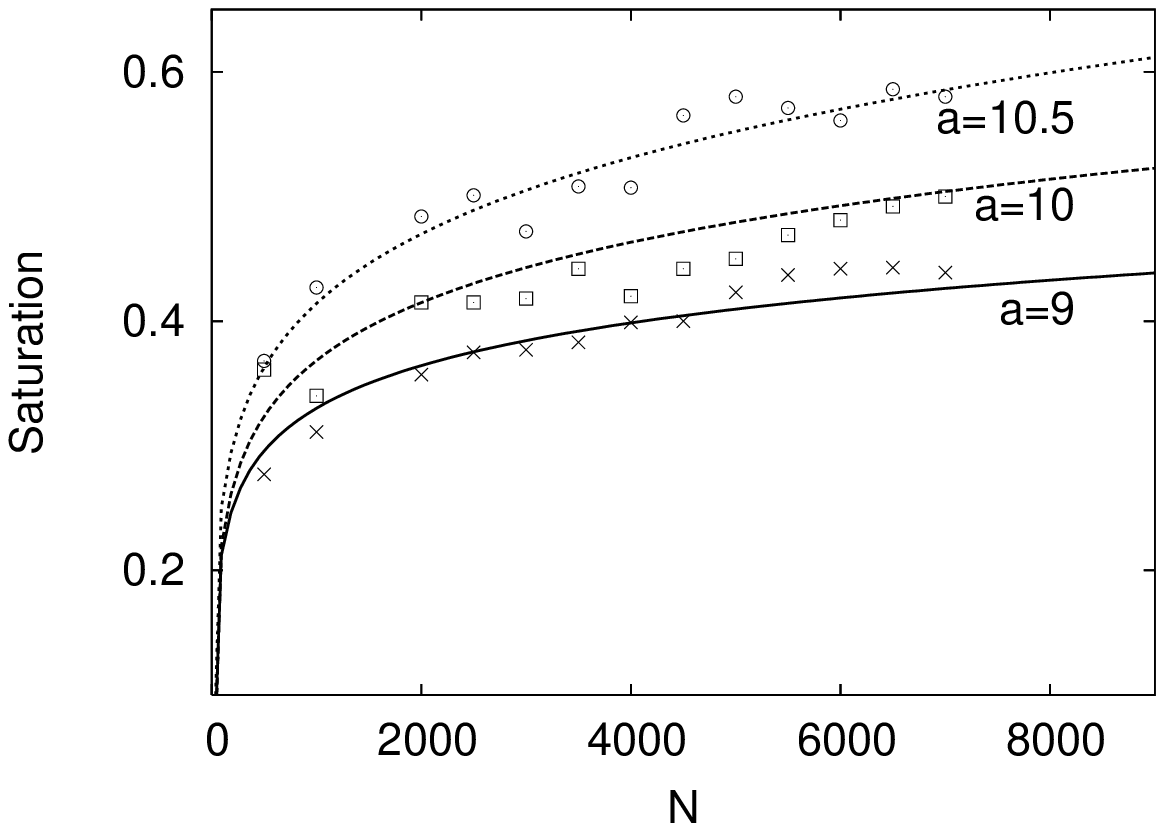}
\includegraphics[width=6cm]{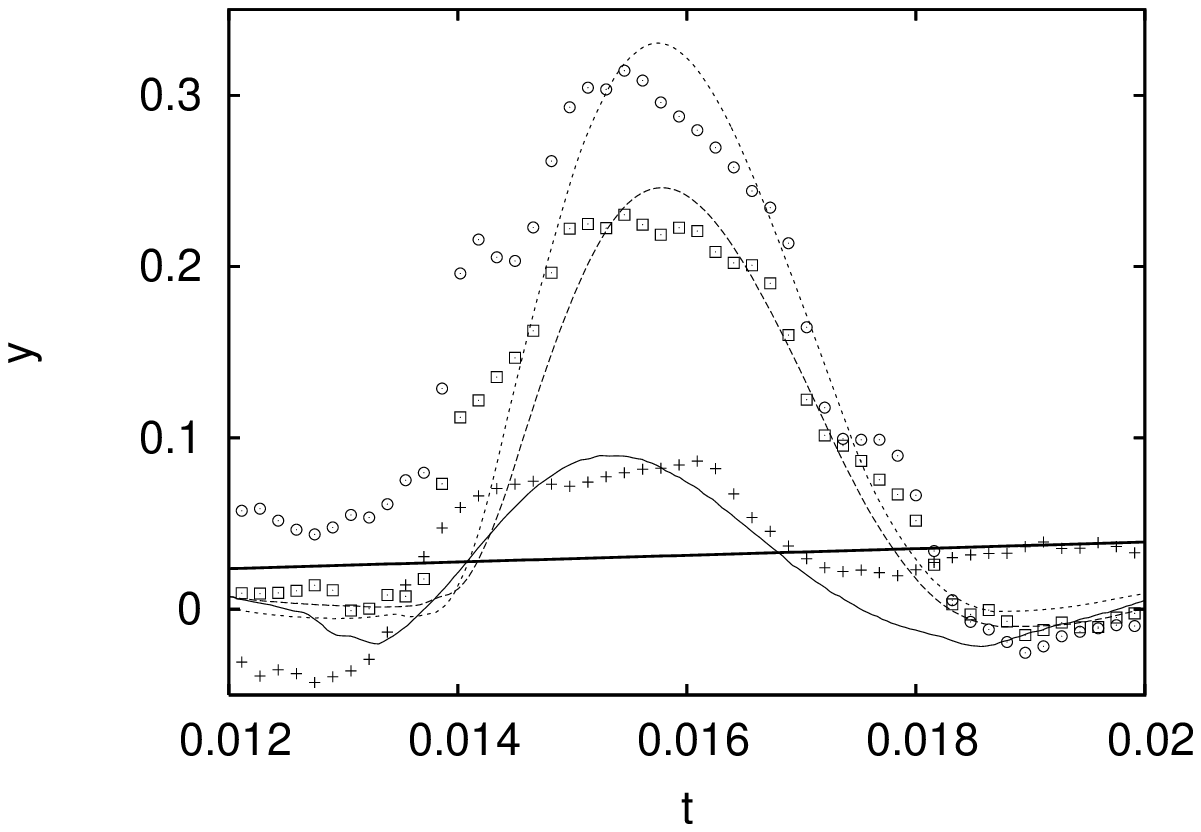}
}
\caption{{\it Left}: Saturation as a function of the energy. {\it Right}: 
Form factor near $\tau_{\rm A}$.
Crosses, squares and circles mark quantum-mechanical results
for $\al=9, 10$ and 10.5, respectively. 
Bold line: GOE result, dashed and dotted lines:
Eq.\ (\ref{sat105}).
\label{fig:satbif}}
\end{figure}

In \fig{fig:satbif} we show the quantum results for $\Delta_\infty$
versus energy ${\tilde E}$  and for the form factor $K(\tau)$ near 
$\tau_{\rm A}$, for the three values $\al=9$, 10 and 10.5
(as crosses, squares and circles, respectively). The solid line gives
the universal GOE prediction, i.e., the first term in \eq{sat10b}. It
agrees well with the quantum result at $\al=9$, in line with
the near chaoticity of the system below the bifurcation. The dashed
and dotted lines show the prediction \eq{sat105}, which includes the 
bifurcating orbits A and L in the uniform approximation, and coincide
well with the quantum results at and above the bifurcation. At the
bifurcation ($\al=\al_4=10$) where $\epsilon=0$, Eq.\ \eq{sat105} is
consistent with the diagonal approximation for the bifurcating orbits 
and thus the same as that used in Ref.\ \cite{ref:BeKeaPra}. 

Figure \ref{fig:satbif} moreover shows that slightly above the bifurcation,
i.e.\ at $\al=10.5$, the additional terms in Eq.\ \eq{sat105}, playing a role
for $\epsilon\neq 0$, give a noticeable contribution, as seen by
the dashed line. The main contribution comes from the term linear in
$\epsilon$ which is the nondiagonal contribution of the pairs of
separate orbits A and L above the bifurcation. To see this, we evaluate
their nondiagonal contribution in the Gutzwiller approximation for
isolated orbits, which would become
\beq
\Delta_{\infty({\rm Gutz)}}^{A.L(\rm non-diag)}=2\frac{A_LA_A}{\pi^2T_A^2}\Big\langle
                 \!\sin\left(\frac{\Delta S}{\hbar}\right)\!\Big\rangle
         \approx \frac{\sqrt{2}}{\pi^2 \epsilon}\frac{\Delta S}{\hbar} 
               = \frac{\epsilon}{2\sqrt{2}\pi^2|a|\hbar}.
\eeq 
(Although the diagonal contribution diverges at the bifurcation, the 
non-diagonal contribution stays finite there.) The additional factor
$1/\sqrt{2}$, compared to the last term in Eq.\ \eq{sat105}, is due to the
fact that the Gutzwiller approximation is not yet valid in this vicinity
of the bifurcation (in particular, the difference in Maslov indices 
is different from the value $1$ reached only far from 
the bifurcation where $\Delta S\gg\hbar$ and hence $\epsilon\gg 1$).

We see therefore that the nondiagonal contribution of the bifurcating
orbits to the saturation value $\Delta_\infty$ is non-negligible in a 
neigborhood above the bifurcation. Note that the value of $\Delta_\infty$ 
is slightly enhanced also by the fact that the particular combination 
of Bessel functions in the uniform approximation \eq{gloapp} can be 
expressed by an Airy funtion (and its derivative, cf.\ Ref.\ \cite{ref:SchoSie}), 
which has its maximum slightly above the bifurcation. This effect is, 
however, not sufficient to explain the enhancement of $\Delta_\infty$ 
found in our results, so that we can argue that the nondiagonal 
contribution is substantial.

It is important to mention that this nondiagonal contribution exists 
as long as $\hbar$ remains finite. In the strict semiclassical limit 
$\hbar\to 0$, the global uniform approximation \eq{gloapp} merges into
the Gutzwiller trace formula for non-zero $\Delta S$, and 
$\sin(\Delta S/\hbar)$ oscillates very fast, so that after the
coarse-graining, the non-diagonal contribution will tend to zero. 
This is expected, since in the semiclassical approximation for mixed 
systems (Eq.\ \ref{sumrule}), periodic orbits with different 
stability give rise to independent statistics.


\section{Conclusions}

In this case study we worked out for the quartic 
oscillator how (pitchfork) bifurcations affect the density of states
and thereby further measures of spectral correlations.
This requires, at a first stage,  detailed knowledge about the
classical bifurcation scenario in that system. At a second
stage, we performed a comprehensive semiclassical calculation
for the density of states invoking uniform approximations for the 
bifurcating orbits involved. 
All features of the coarse-grained quantum density of states
are adequately, and to high precision (mean level spacing), 
semiclassically reproduced, which is not evident in such a
system with mixed phase space dynamics. Our semiclassical 
evaluation of the spectral rigidity close to the bifurcation shows 
strong deviations from the RMT behaviour, even though the phase
space is predominantly chaotic and the bifurcation-affected
phase space region appears negligible. This confirms
that spectral statistics is rather susceptible with respect to bifurcation
effects. Moreover we could unreval the role of orbit pairs
born at the bifurcation which prevail with near-degenerate
actions for larger control parameter regimes and strongly affect
the spectral statistics. Such orbit pairs are obviously 
classically correlated and require a treatment beyond the
diagonal approximation. 

This analysis moreover implies that in a comprehensive semiclassical 
approach to spectral correlations in mixed systems, which still 
remains as a challenge, off-diagonal contributions in the occuring 
multiple sums over periodic orbits should be considered, analogously 
to the purely hyperbolic case.

Further open questions not answered in the present work 
include a corresponding analysis of how eigenstates are affected 
at a bifurcation. Finally, studies of bifurcation signatures
in other observables such as quantum transport are still rare
\cite{ref:KeaPraSie} and remain to be explored.

\section{Acknowledgments}
We thank S. Creagh, J. Keating, P. Schlagheck and M. Sieber 
for useful discussions and are grateful to K. Jänich for his
assistance in evaluating the boundary term in Eq.\ \eq{denQO}.
The numerical determination of the periodic orbits and their
stabilities was done with the program developed by Ch. Amann
in \cite{ref:amann}. We acknowledge financial support of the 
{\em Deutsche Forschungsgemeinschaft} (GRK 638). We are grateful
to B. Zhilinskii for fruitful criticism and to our anonymous 
referees for a number of valuable corrections,



\begin{thebibliography}{00}
\bibitem{footnote1}
The parameter $a$ appearing in the corresponding
normal form, in the notation of \cite{ref:SchoSie}, was obtained
analytically in \cite{ref:BraFeMagMeh}.
\end{thebibliography}
\end{document}